\documentclass[letterpaper,11pt]{article}
\usepackage[DIV12]{typearea}
\usepackage[utf8]{inputenc}
\usepackage[T1]{fontenc}
\usepackage{lmodern}
\usepackage{amsmath,amssymb,mathtools,booktabs,longtable,array,graphicx,float,xspace}
\usepackage{dsfont}
\usepackage{pdflscape,cite,braket}
\usepackage[hidelinks]{hyperref}
\usepackage{cleveref}
\usepackage{yhmath}
\usepackage[nolist,nohyperlinks]{acronym}
 \usepackage[normalem]{ulem}
 
\renewcommand{\baselinestretch}{1.2}
\DeclareMathOperator{\Ab}{Ab}
\DeclareMathOperator{\Out}{Out}
\DeclareMathOperator{\Aff}{Aff}
\DeclareMathOperator{\Hom}{Hom}

\newcommand{\Z}[1]{\ensuremath{\mathds{Z}_{#1}}}

\newcommand{\SU}[1]{\ensuremath{\mathrm{SU}(#1)}}
\newcommand{\SL}[1]{\ensuremath{\mathrm{SL}(#1)}}
\newcommand{\GL}[1]{\ensuremath{\mathrm{GL}(#1)}}
\newcommand{\U}[1]{\ensuremath{\mathrm{U}(#1)}}

\newcommand{\e}{\mathrm{e}}
\newcommand{\I}{\mathrm{i}}
\newcommand{\Id}{\mathds{1}}

\newcommand{\CP}{\ensuremath{\mathcal{CP}}\xspace}
\newcommand{\x}{\ensuremath{\times}}
\newcommand{\dd}{\mathrm{d}}

\newcommand{\ZZ}{\mathds{Z}}
\newcommand{\RR}{\mathds{R}}
\newcommand{\Abhat}[1]{\widehat{\Ab}(#1)}
\newcommand{\Gshift}{\mathcal G_{\rm shift}}
\newcommand{\Ggeo}{\mathcal G_{\rm geo}}
\newcommand{\Outgeo}{\Out_{\rm geo}}
\newcommand{\Outlin}{\Out_{\rm lin}^{\rm geo}}

\usepackage{xcolor}

\definecolor{redM}{RGB}{206, 0, 0}

\newsavebox{\tabstackbox}
\newcommand{\tstack}[1]{%
  \sbox{\tabstackbox}{\begin{tabular}[t]{@{}c@{}}#1\end{tabular}}%
  \ifdim\wd\tabstackbox>\linewidth
    \resizebox{\linewidth}{!}{\usebox{\tabstackbox}}%
  \else\usebox{\tabstackbox}\fi}
\hypersetup{pdftitle={The flavor of non-Abelian orbifolds},
pdfauthor={Hernandez-Segura, Liu, Perez-Martinez, Ramos-Sanchez}}

\everydisplay\expandafter{\the\everydisplay\thickmuskip=11mu plus 5mu}
\makeatletter
\g@addto@macro\bfseries{\boldmath}
\makeatother

\begin{document}
\renewcommand{\thefootnote}{\alph{footnote}}
\hypersetup{pageanchor=false}
\begin{titlepage}

\vspace*{1.0cm}

\begin{center}
{\Large\textbf{\boldmath The Flavor of Non-Abelian Orbifolds\unboldmath}}

\renewcommand*{\thefootnote}{\fnsymbol{footnote}}

\vspace{1cm}
\textbf{Miguel Hern\'andez--Segura$^{a,}$\footnote{\texttt{hernandez@estudiantes.fisica.unam.mx}}}, 
\textbf{Xiang-Gan Liu$^{a,}$\footnote{\texttt{xgliu@fisica.unam.mx}}}, 
\textbf{Ricardo P\'erez--Mart\'inez$^{b,}$\footnote{\texttt{ricardo.perezmartinez@uadec.edu.mx}}},\\
and
\textbf{Sa\'ul Ramos--S\'anchez$^{a,}$\footnote{\texttt{ramos@fisica.unam.mx}}}\\[5mm]
\textit{$^a$\small Instituto de F\'isica, Universidad Nacional Aut\'onoma de M\'exico, Cd.\ de M\'exico C.P.\ 04510, M\'exico}
\\[5mm]

\textit{$^b$\small Facultad de Ciencias F\'isico-Matem\'aticas, Universidad Aut\'onoma de Coahuila, Edificio A, Unidad Camporredondo, 25000, Saltillo, Coahuila, M\'exico}

\end{center}
\vspace{1cm}
\vspace*{1.0cm}

\begin{abstract}
In string model building, flavor symmetries can arise from the geometric features of the compactification space. In the case of heterotic orbifold models, modular and traditional flavor symmetries can be related to both geometric and algebraic properties of the orbifold space group. Our goal is to study the origin of moduli-independent discrete flavor symmetries in symmetric heterotic orbifolds with non-Abelian point groups. We develop various methods based on the Abelianization and the geometric automorphisms of the orbifold space group
to determine these symmetries and their associated charges for matter fields. Notably, our methods were applied to all 331 non-Abelian affine 
geometries, including roto-translations, suitable for delivering $\mathcal N=1$ supersymmetric heterotic compactifications in four dimensions. This provides additional motivation to pursue the derivation of new types of phenomenologically viable models from string theory.
\end{abstract}
\end{titlepage}

\hypersetup{pageanchor=true}
\setcounter{footnote}{0}
\renewcommand{\thefootnote}{\arabic{footnote}}

{\small\renewcommand{\baselinestretch}{1}\selectfont\setlength{\parskip}{0pt}\tableofcontents}

\clearpage
\section{Introduction}
\label{sec:introduction}
The origin of fermion mass hierarchies and mixing remains an open problem in
particle physics. Discrete flavor symmetries can constrain Yukawa couplings
\cite{Altarelli:2010gt,Ishimori:2010au}, but in bottom-up models the symmetry
group and matter representations are usually assumed. Heterotic orbifolds~\cite{Bailin:1999nk,Ramos-Sanchez:2008nwx,Vaudrevange:2008sm,Ramos-Sanchez:2024keh}
provide a framework in which both arise from the compactification geometry
\cite{Nilles:2012cy,Nilles:2014owa}: twisted strings are localized at fixed points or fixed
tori, their interactions obey space-group selection rules
\cite{Hamidi:1986vh,Dixon:1986qv}, and equivalent localizations can be related
by geometric symmetries. These geometric selection rules are based on the study of  couplings among twisted string states in \ac{CFT}~\cite{Dixon:1985jw,Dixon:1986jc,Lauer:1989ax,Lauer:1990tm}. Beyond orbifolds, discrete and modular flavor symmetries have been derived in heterotic Calabi--Yau and magnetized compactifications, including geometric automorphisms, Yukawa couplings, and selection rules~\cite{Cremades:2004wa,Buchbinder:2016yuh,Candelas:2017ive,Lukas:2017vqp,Braun:2017cys,Gray:2021fcq,Ishiguro:2021ccl,Almumin:2021fbk,Ishiguro:2024mhy,Kobayashi:2024fsm,Dong:2025csr,Dong:2026yte}. Also geometric constructions of non-Abelian flavor symmetries and their heterotic realizations have been investigated extensively~\cite{Adulpravitchai:2009xyz,Carballo-Perez:2016ooy,Beye:2014nxa}.

For orbifolds with Abelian point groups, the methods to determine traditional flavor symmetries are well understood. Permutations of
equivalent localizations combine with selection-rule phases to generate
non-Abelian groups such as $D_8$ and $\Delta(54)$
\cite{Kobayashi:2004ya,Ko:2007dz,Kobayashi:2006wq}. This
framework has been studied systematically, including nonfactorizable
lattices and roto-translations \cite{Olguin-Trejo:2018wpw}, with the phase
symmetry being encoded by the space-group Abelianization
\cite{Ramos-Sanchez:2018edc}. In the Narain description these transformations
sit inside the broader structure of target-space dualities, modular 
symmetries and \CP 
\cite{Narain:1985jj,Narain:1986am,GrootNibbelink:2017usl,Baur:2019kwi,Baur:2019iai,Nilles:2020gvu,Baur:2020jwc,Baur:2024qzo}. Complementary perspectives emerge from recent analyses of twisted-sector modular representations and subsequent string-derived modular flavor symmetries~\cite{Kobayashi:2018rad,Nilles:2020tdp,Nilles:2020nnc,Baur:2021mtl,Li:2025bsr}, while
related Narain-based constructions extend to asymmetric non-geometric orbifolds~\cite{Baykara:2024vss} and
non-supersymmetric heterotic compactifications~\cite{Funakoshi:2025lxs}.

However, for non-Abelian point groups the extension of the mentioned methods is not automatic. Twisted sectors
are labeled by point-group conjugacy classes, which exhibit a richer structure 
than in Abelian orbifolds.
Hence, in this case, neither the point
group nor the number of fixed components determines the flavor structure. 
In this work, we focus on the 331 six-dimensional non-Abelian affine geometries that are compatible with four-dimensional $\mathcal N=1$ supersymmetry~\cite{Fischer:2012qj,Fischer:2013qza}. While previous research addressed the spectra and gauge-group properties of specific models within these classified geometries~\cite{Konopka:2012gy,Fischer:2013qza,Hernandez-Segura:2025sfr}, the determination of their complete traditional flavor groups is the primary subject of the present analysis.

It is worth to mention that recent studies have investigated conjugacy-class selection rules in
non-Abelian string orbifolds~\cite{Kaidi:2024sgr,Kobayashi:2025nis};
and related non-invertible flavor structures have been analyzed in intersecting
and magnetized D-brane models~\cite{Funakoshi:2024nif}. Nevertheless, such selection rules 
are distinct from the conventional group actions studied here, which determine 
moduli-independent geometric flavor symmetries.

The goal of this work is also to develop a complete formal geometric framework for deriving the moduli-independent flavor symmetries of arbitrary orbifolds defined by a space group $S$. A first component of this symmetry is the Abelianization group $\Ab(S)=S/[S,S]$ associated with the selection rules governing couplings among string matter states~\cite{Ramos-Sanchez:2018edc} with their charges given by the phases of the character group 
$\Abhat{S}=\Hom(\Ab(S),\U1)$. One of the challenges that has not been properly addressed yet is the effect on the symmetries and charges of the roto-translations in $S$. A second ingredient for moduli-independent flavor symmetries has naively been taken to be the group of permutations of localized states in the orbifold~\cite{Kobayashi:2006wq,Olguin-Trejo:2018wpw}. In fact, as we shall see, one must rather derive the actual geometric permutations from affine transformations that normalize $S$ and, simultaneously, preserve the metric and antisymmetric tensor. This will also have an effect on the charges. A third and occasionally ignored subgroup of moduli-independent flavor symmetries is the set of space-group outer automorphisms associated with target-space modular transformations that leave the moduli untouched, such as $\mathsf S^2=-\Id$, which can also be associated with string selection rules~\cite{Knapp-Perez:2025cht,Liu:2025lym}. These elements are combined together in a multiplicative closure, which builds a non-Abelian traditional flavor group. This work provides the details of this construction, which can be achieved without the use of a separate \ac{CFT} calculation.

Our paper is organized as follows. In \Cref{sec:overview}, we present a brief introduction to the formalism of heterotic orbifolds, which allows us to fix our notation. \Cref{sec:abelianization} is then devoted to developing the techniques to arrive at the Abelianization symmetry, including the effects of roto-translations and the charges especially for non-Abelian orbifolds. In \Cref{sec:geometric}, we determine the moduli-independent transformations that include affine translations, related to permutations of matter states, and linear geometric symmetries. All these elements are put together in \Cref{sec:flavorgroups} to build the general form of the flavor symmetry groups. In \Cref{sec:algorithm}, we describe the algorithm that we follow to arrive at the application of our methods to the 331 non-Abelian affine geometries for heterotic orbifolds. Our conclusions are given in \Cref{sec:conclusions}. Appendices are devoted to details of our findings; in particular, \Cref{app:abelian_catalogue,app:fullcatalogue,app:imagegroups} provide the details of our full catalogue of charges and symmetries of non-Abelian orbifolds.

\section{Heterotic orbifolds and geometric moduli}
\label{sec:overview}

A six-dimensional symmetric toroidal orbifold is defined by the quotient
\begin{equation}
\mathcal O:=\RR^6/S\,,
\label{eq:orb}
\end{equation}
where $S$ is the space group identified as a group extension of the point group $P$ by the translation lattice $\Lambda$. These elements satisfy in general the short exact sequence
\begin{equation}
1\longrightarrow\Lambda\,\longrightarrow\, S\,\longrightarrow\,P\longrightarrow 1\,,
\label{eq:ext_space}    
\end{equation}
which implies that $P\cong S/\Lambda$.  When the sequence splits, $S=\Lambda \rtimes P$. Otherwise, point-group elements given as pure rotations do not appear in $S$, and there are space group elements called {\it roto-translations} that combine point group elements with (fractional) translations not included in $\Lambda$. We consider the 331 non-Abelian geometries of Refs.~\cite{Fischer:2012qj,Fischer:2013qza}, with $P\subset \SU3$ in an appropriate complex basis.\footnote{At the level of this work, gauge embeddings and Wilson lines appearing in heterotic orbifolds are not relevant as we only focus on the consequences of the geometric input.}

Let $e=(e_1,\ldots,e_6)$ be the geometric vielbein whose columns span the torus lattice,
\begin{equation}
\Lambda=\{e\,n\mid n\in\ZZ^6\}.
\label{eq:lattice}
\end{equation}
Coordinates in the covering space are denoted by $y\in\RR^6$.  We use dimensionless lattice coordinates
\begin{equation}
y=e\,\widehat y \,.
\label{eq:basischange}
\end{equation}
Accordingly, a point-group element $\vartheta\in P\subset O(6)$ in the coordinate basis is
represented in the lattice basis by
\begin{equation}
\widehat\vartheta=e^{-1}\vartheta e\in \GL{6,\ZZ}\,.
\label{eq:thetahat}
\end{equation}
We denote the corresponding lattice realization of the point group by
$\widehat P=e^{-1}Pe\subset \GL{6,\ZZ}$.
A general space-group element is written in the coordinate basis as
\begin{equation}
g=(\vartheta,\mu_\vartheta+e\,n)\,,
\qquad
\mu_\vartheta=e\,\widehat\mu_\vartheta\,,
\qquad
n\in\ZZ^6 \,,
\label{eq:gcoord}
\end{equation}
and acts as $y\mapsto \vartheta y+\mu_\vartheta+e\,n$.
Conjugating by the vielbein changes to the lattice basis,
\begin{equation}
\widehat g
:=(e^{-1},0)\,g\,(e,0)
=
(\widehat\vartheta,\widehat\mu_\vartheta+n)\,.
\label{eq:ghat}
\end{equation}
A nonintegral vector $\widehat\mu_\vartheta$ describes a fractional translation, which is essential to define a roto-translation.
Likewise, we write a general geometric affine transformation as
\begin{equation}
\alpha=(\sigma,e\,t)
\qquad\Longleftrightarrow\qquad
\widehat\alpha=(\widehat\sigma,t),
\qquad\text{with}\quad
\widehat\sigma=e^{-1}\sigma e 
\quad\text{and}\quad
t\in\RR^6\,.
\label{eq:alphabasis}
\end{equation}
Thus, $\sigma$ acts on the physical coordinates $y$, whereas $\widehat\sigma$ is the matrix
used in the crystallographic computation.  Following the conventions of
Refs.~\cite{Baur:2020jwc,Nilles:2020gvu}, the translation vectors $n$ and $t$ are already defined in the lattice basis and are left unhatted. 
The multiplication and inverse are
\begin{equation}
(\vartheta,v)(\vartheta',v')=(\vartheta\vartheta',v+\vartheta v')\,,
\qquad 
(\vartheta,v)^{-1}=(\vartheta^{-1},-\vartheta^{-1}v)\,.
\label{eq:mrS}
\end{equation}

A comment on the fractional shifts $\mu_\vartheta$ is in order. 
In some cases, fractional translations can be eliminated by a suitable shift of the origin or a different choice of space-group generators. A space group $S$ is called {\it symmorphic} if all translations can be made in $\Lambda$, such that the sequence~\eqref{eq:ext_space} splits and $S=\Lambda\rtimes P$. Otherwise, $S$ is called {\it non-symmorphic} and includes roto-translations. Notice that a fractional shift is therefore not, by itself, enough to establish that $S$ is non-symmorphic. However, in this work we shall assume that all space groups that we consider are in bases such that fractional translations only appear in roto-translations within non-symmorphic space groups.

In heterotic orbifolds, string states are defined by the conditions ensuring that the string closes after one worldsheet period. In terms of space-group elements, these conditions read
\begin{equation}
y(\sigma+2\pi)=\vartheta y(\sigma)+v\,,
\qquad g=(\vartheta,v)\in S\,,
\label{eq:boundarycondition}
\end{equation}
where $g$ is called a {\it constructing element}. Untwisted strings correspond to constructing elements with $\vartheta=\Id$ whereas $\vartheta\neq\Id$ leads to twisted strings (i.e.\ closing only due to the twist). Such twisted strings are localized at fixed loci $y_f$ satisfying
\begin{equation}
(\Id-\vartheta)y_f=v\,.
\label{eq:fpeq}
\end{equation}
When $\Id-\vartheta$ is invertible, $y_f$ corresponds to an isolated fixed point.
Otherwise, the condition~\eqref{eq:fpeq} can be inconsistent or have a family of solutions
along $\ker(\Id-\vartheta)$. After identifying points related by lattice translations, the latter becomes a fixed torus in supersymmetric models. Different constructing elements can describe the same boundary condition. In particular, if they belong to the same conjugacy class, $[g]=\{hgh^{-1}\mid h\in S\}$, they are physically equivalent.  For a given point-group twist $\vartheta$, one must identify elements that are related by 
space-group conjugation. 

Traditionally, we collect the twisted strings in various twisted sectors, which are labeled by conjugacy classes $[\vartheta]$ of the point group $P$. For each sector, the corresponding constructing elements determine the inequivalent localized fixed points or fixed tori. We denote the set of these localization labels by $\mathcal L_{[\vartheta]}$, their union over the localized twisted sectors by $\mathcal L$ and their transformation by $\rho_{\mathcal L}$. These geometric labels should be distinguished from the physical string states, which also depend on the gauge embedding and the orbifold projection.

\subsection{Action on the geometric background}
The metric $G$ and antisymmetric $B$ tensor are most conveniently written in the lattice basis.  With
$G=e^Te$, the relevant sigma-model term can be written schematically as
\begin{equation}
S_\sigma\supset\frac{1}{4\pi\alpha'}\int \dd^2z\,
(G_{ij}+B_{ij})\,\partial\widehat y^i\bar\partial\widehat y^j \,.
\label{eq:sigma}
\end{equation}
Under the active affine transformation
\begin{equation}
\widehat y\longmapsto \widehat\sigma\,\widehat y+t
\qquad\Longleftrightarrow\qquad
y\longmapsto \sigma y+e\,t \,,
\label{eq:affineaction}
\end{equation}
the constant translation drops out of the sigma model, while the two tensors transform as
\begin{equation}
G\longmapsto \widehat\sigma^T G\widehat\sigma \,,
\qquad
B\longmapsto \widehat\sigma^T B\widehat\sigma \,.
\label{eq:GBtransform}
\end{equation}
Hence, a geometric transformation is unbroken at a given background precisely when it leaves
$G$ and $B$ invariant.
In the more frequent (passive) convention used in the Narain literature, the corresponding transformations are
\begin{equation}
e\longmapsto e\,\widehat\sigma^{-1}\,,
\qquad
G\longmapsto \widehat\sigma^{-T}G\widehat\sigma^{-1}\,,
\qquad
B\longmapsto \widehat\sigma^{-T}B\widehat\sigma^{-1}\,.
\label{eq:passiveGB}
\end{equation}
The two descriptions are equivalent, and the fixed-background condition is the same in either
convention.

Consistency demands that the orbifold point group itself preserve the background.  Therefore, the allowed metric and
$B$-field moduli must satisfy
\begin{equation}
\widehat\vartheta^T G\widehat\vartheta\stackrel!=G
\qquad\text{and}\qquad
\widehat\vartheta^T B\widehat\vartheta\stackrel!=B
\qquad\text{with}\quad
\widehat\vartheta\in\widehat P
\label{eq:GBinvariance}
\end{equation}
and $G=G^T>0$ and $B=-B^T$. Choosing bases of these solution spaces yields the parametrizations
\begin{equation}
G(\mathbf g)=\sum_{i=1}^{n_G}g_iG_i\,,
\qquad
B(\mathbf b)=\sum_{a=1}^{n_B}b_aB_a \,.
\label{eq:GBparam}
\end{equation}
For a non-Abelian point group, one imposes the conditions~\eqref{eq:GBinvariance} simultaneously for all
generators.  Since the roto-translations $\widehat\mu_\vartheta$ do not enter these equations,
all affine classes belonging to the same linear $\mathbb Z$-class have the same $G,B$
parametrization.
Moreover, if the character associated with $\widehat\vartheta$ is $\chi(\widehat\vartheta):=\mathrm{tr}\,\widehat\vartheta$ in the real
six-dimensional representation, then the numbers $n_G$ of symmetric and $n_B$ of antisymmetric moduli allowed by $P$ are given by
\begin{equation}
n_G=\frac1{|P|}\sum_{\vartheta\in P}\frac{\chi(\widehat\vartheta)^2+\chi(\widehat\vartheta^2)}{2}\,,
\qquad
n_B=\frac1{|P|}\sum_{\vartheta\in P}\frac{\chi(\widehat\vartheta)^2-\chi(\widehat\vartheta^2)}{2}\,.
\label{eq:charcheck}
\end{equation}

\subsection{Example: \texorpdfstring{$T^2/\Z3$}{T2/Z3}}
Let us briefly explore the introduced elements in a simple Abelian example. For the $T^2/\Z3$ orbifold, we choose the twist in the lattice basis as
\begin{equation}
\widehat\theta=
\begin{pmatrix}0&-1\\1&-1\end{pmatrix},
\qquad\text{satisfying}\qquad
\theta=e\,\widehat\theta\,e^{-1}\,,
\qquad
\widehat\theta^3=\Id_2 \,.
\end{equation}
The solutions to the conditions~\eqref{eq:GBinvariance} are
\begin{equation}
G=R^2
\begin{pmatrix}1&-\tfrac12\\-\tfrac12&1\end{pmatrix},
\qquad
B=\beta
\begin{pmatrix}0&1\\-1&0\end{pmatrix}.
\label{eq:Z3GB}
\end{equation}
We see that, up to the size $R$, the metric $G$ of the two-dimensional space of this orbifold corresponds to the Cartan matrix of the \SU3 Lie algebra.
The $180^\circ$ rotation $C=-\Id_2$ preserves both the geometry and the $B$-field of the orbifold for arbitrary reals $R$ and
$\beta$, so it is an orbifold-compatible, unbroken transformation that survives everywhere in the allowed moduli space.  A reflection $\begin{pmatrix}0&1\\1&0\end{pmatrix}$ can preserve the (hexagonal) metric $G$ while reversing the $B$-field and therefore survives only on a \CP-invariant
sublocus.  This is the basic distinction between generic traditional flavor transformations and
modular/\CP enhancements \cite{Baur:2019kwi,Baur:2019iai,Nilles:2020gvu}.

\section{Space-group Abelianization and discrete charges}
\label{sec:abelianization}

\subsection{Space-group selection rule and character symmetry}

Let $g_a$, $a=1,\ldots,m$, be constructing elements of localized twisted
strings. The space-group selection rule on the sphere requires representatives
of their conjugacy classes whose ordered product is the identity $\Id_S=(\Id,0)$
\cite{Hamidi:1986vh,Dixon:1986qv,Ramos-Sanchez:2018edc}:
\begin{equation}
\exists\ h_a\in S\,,
\qquad\text{such that}\qquad
\prod_{a=1}^{m}h_ag_ah_a^{-1}=\Id_S\,.
\label{eq:sgsr}
\end{equation}
Besides this necessary condition for a coupling, other selection rules, such as gauge symmetry, must also be imposed~\cite{Kobayashi:2004ya,Kobayashi:2011cw,Nilles:2013lda}. 
To study the consequent family symmetry associated with the space-group selection rule, consider an Abelian homomorphism $s$ that is constant on conjugacy classes of $S$, i.e.~\cite{Ramos-Sanchez:2018edc}
\begin{equation}
s(hgh^{-1})=s(g)\,,\qquad 
s(gh)=s(g)s(h)\,,\qquad 
g,h\in S\,.
\label{eq:sproperties}
\end{equation}
The first condition ensures that the charge does not depend on the chosen
representative of a twisted string; the second ensures that charges combine
consistently in an interaction. Together, they lead to
\begin{equation}
s(h)s(g)s(h)^{-1}=s(g)\quad\implies\quad
s([g,h])=s(g)s(h)s(g)^{-1}s(h)^{-1}=1=s(\Id_S)\,,
\label{eq:commutatorphase}
\end{equation}
which stresses that $s$ loses the non-Abelian properties of $S$ because all images commute.
Further, this implies that every commutator carries a trivial charge.
The most general charge assignment is thus obtained by identifying
elements that differ by a product of commutators, namely by passing to
the {\it Abelianization} of the space group $\Ab(S)=S/[S,S]$. These observations add up to state that the projection $s:S\to\Ab(S)$ encodes the Abelian symmetries associated with the space-group selection rule.\footnote{Any map satisfying
\Cref{eq:sproperties} factors through this projection. A particular
phase representation can identify further elements, so its kernel need
not be exactly $[S,S]$.} The properties~\eqref{eq:sproperties} of the projection $s$ can be also written in its additive form $q$ characterized by
\begin{equation}
    q(hgh^{-1})=q(g)\,,\qquad 
    q(gh)=q(g)+q(h)\,,\qquad 
    q([g,h])=0\,.
\end{equation}
In these terms, applying the projection to \Cref{eq:sgsr} yields\footnote{The multiplicative form of the projection $s$ acting on \Cref{eq:sgsr} is $\prod_{a=1}^{m} s(g_a) = s(\Id_S) = 1$.} 
\begin{equation}
  \sum_{a=1}^{m}q(g_a)=0\,.
\label{eq:abelian_selection}
\end{equation}
Note that the Abelianization group $\Ab(S)$ should always be decomposed into cyclic factors. Once this happens, $q(g)$ corresponds to the usual discrete charges modulo their respective orders.
For a non-Abelian space group the converse need not hold. The Abelianization retains the additive selection rule~\eqref{eq:abelian_selection} in terms of the charges $q(g)$, not the full space-group selection rule.

Let us now proceed to determine the Abelianization charges from the space-group relations and then analyze their properties in the context of non-Abelian orbifolds.

\subsection{Flavor symmetries from space-group presentations}
\label{sec:presentation}

Let $\vartheta_a$, $a=1,\ldots,r$, generate $P$, and consider for the general case of non-symmorphic space groups lifts
$g_a=(\vartheta_a,\mu_a)$ together with the generators of lattice translations
$T_i=(\Id,e_i)$, $i=1,\dots,6$. The space-group presentation is defined in general by the relations
\begin{equation}
[T_i,T_j]=(\Id,0)=\Id_S,\qquad
g_aT_i g_a^{-1}=\prod_jT_j^{(\widehat\vartheta_a)_{ji}},
\qquad R_b(g_1,\ldots,g_r)\Bigl(\prod_iT_i^{-c_{bi}}\Bigr)=\Id_S,
\label{eq:sgpresentation}
\end{equation}
where $c_{bi}\in\ZZ$ specifies the lattice translation
generated by the $b$-th point-group relation after lifting
its generators to the space group; explicitly, $R_b(g_1,\ldots,g_r)=(\Id,\sum_i c_{bi}e_i)$. Here, $R_b=\Id$ are the defining relations of the point group, whereas their lifts in $S$ may deliver nontrivial lattice translations.\footnote{Note in \Cref{eq:sgpresentation} that the relations $R_b$ are applied to elements of $S$, which imply non-Abelian constraints on the whole space group, not only on $P$, arising from the presentation of $P$.}
Substituting the affine generators into such a relation can leave a nonzero lattice
translation $e\,c_b$ in the presence of roto-translations. For example, if
$\vartheta_a$ has order $N_a$, then
\begin{equation}
g_a^{N_a}=(\Id,\lambda_a),\qquad
\lambda_a=\sum_{j=0}^{N_a-1}\vartheta_a^j\mu_a\in\Lambda.
\label{eq:liftpower}
\end{equation}

In the Abelianization, let us simplify the notation by setting $u_a:=q(g_a)$ and $v_i:=q(T_i)$. If $r_{ba}$ is the sum of all exponents of $g_a$ appearing in $R_b$, then \Cref{eq:sgpresentation} implies
\begin{equation}
v_i-\sum_j(\widehat\vartheta_a)_{ji}v_j=0,
\qquad
\sum_a r_{ba}u_a-\sum_i c_{bi}v_i=0.
\label{eq:Abrelations}
\end{equation}
These are ordinary linear relations between discrete charges. Arranging
their integer coefficients in a matrix and reducing it to Smith normal
form gives the independent cyclic factors of $\Ab(S)$
\cite{Ratcliffe:2009,Ramos-Sanchez:2018edc}.\footnote{Smith normal form uses
integer row and column operations with integer inverses to diagonalize
the relation matrix. A nonzero diagonal entry $d>1$ gives a factor
$\ZZ_d$; an entry one removes a generator. Unconstrained generators would
give free $\ZZ$ factors. The accompanying change of basis determines the
charges as integer combinations of the original exponents.}
For example, the Abelianization of the second relation in \Cref{eq:sgpresentation} reads
\begin{equation}
s(T_i)=\prod_j s(T_j)^{(\widehat\vartheta_a)_{ji}}\,.
\label{eq:translationidentification}
\end{equation}
More explicitly, since $g_aT_ig_a^{-1}=(\Id,\vartheta_a e_i)$, \Cref{eq:translationidentification} can be rewritten as
\begin{equation}
    s(\Id,e_i) = s(\Id,\vartheta_a e_i)
    \qquad\implies\qquad
    s(\Id,\lambda)=s(\Id,\vartheta_a\lambda)
\end{equation}
where $\lambda = \sum_in_i e_i \in\Lambda$.
This means that lattice translations related by a point-group rotation carry the same Abelianization charge. Hence, if
$\lambda+\vartheta\lambda+\cdots+\vartheta^{m-1}\lambda=0$, then
\begin{equation}
\label{eq:translationorderbound}
   s\bigl((\Id,\lambda)\bigr)^m=1\,.
\end{equation}
This provides an upper bound on the order of the translation charge. Its
actual order, and its independence from the other charges, follow only
after all relations in \Cref{eq:Abrelations} have been imposed.

The translation relations can be collected in the sublattice
\begin{equation}
\widehat\Lambda_P=\sum_{\widehat\vartheta\in\widehat P}
(\Id-\widehat\vartheta)\ZZ^6,
\qquad \Lambda_P=e\widehat\Lambda_P.
\label{eq:KP}
\end{equation}
It contains the translation differences that have become trivial charges
and depends only on the linear point-group action.\footnote{The quotient
$\Lambda/\Lambda_P$ is often called the coinvariant group. It identifies
$\lambda$ with $\vartheta\lambda$ for every $\vartheta\in P$. This is
standard group-homology terminology; see e.g.\
Ref.~\cite{Brown:1982}. In the present work, the group is finite.}
If the space group has no roto-translations, i.e.\ if $S=P\ltimes\Lambda$, the
point-group and translation relations separate. One finds
\begin{equation}
[S,S]=\Lambda_P\rtimes[P,P],\qquad
\Ab(S)\cong\frac{\widehat\Lambda}{\widehat\Lambda_P}\x\frac{P}{[P,P]}
=\frac{\ZZ^6}{\widehat\Lambda_P}\x\Ab(P)\,.
\label{eq:Absplit}
\end{equation}
The first factor contains the surviving translation charges, and the
second contains the point-group charges. The first factor is finite here
because the point group has no common invariant direction.\footnote{In
particular, $\sum_{\vartheta\in P}\vartheta\lambda=0$ for every
$\lambda\in\Lambda$. Since all terms have the same class in
$\Lambda/\Lambda_P$, each class is annihilated by $|P|$.}
The point-group Abelianizations are listed in
Appendix~\ref{app:pointgroups}.

\subsection{Conservation of discrete charges}
\label{sec:chargeconservation}
Once the method to determine the Abelianization charges and structure has been established, we can now proceed to figure out the corresponding phase transformations. A general space-group element can be
written as $g=T_1^{n_1}\cdots T_6^{n_6}w(g_1,\ldots,g_r)$, where $w$ is
a word in the chosen point-group lifts. If $k_a$ is the signed sum of all exponents of $g_a$ appearing in such a word, then we find the equivalent Abelianization equations
\begin{equation}
s(g)=\prod_{i=1}^{6}s(T_i)^{n_i}\prod_{a=1}^{r}s(g_a)^{k_a}
\qquad\text{and}\qquad
q(g)=\sum_i n_i v_i+\sum_a k_a u_a\,.
\label{eq:chargefromword}
\end{equation}
Different words representing the same element give the same charge
because they are related by \Cref{eq:Abrelations}. No special normal
form for a general non-Abelian point group is required.

The phase transformations associated with these charges build the group\footnote{Prior to this expression, hats were used only on geometric quantities to indicate their 
lattice-basis representations. In contrast, the hat on $\widehat{\Ab}(S)$ denotes the character group $\Hom(\Ab(S),\U1)$. Hence, the two uses of hats, despite being standard, are unrelated.}
\begin{equation}
\Abhat{S}:=\Hom(\Ab(S),\U1).
\label{eq:Abhat}
\end{equation}
If $\Ab(S)\cong\prod_A\Z{N_A}$, a character labeled by
$c_A\in\Z{N_A}$ acts as
\begin{equation}
\chi_c(q):=\exp\!\left(2\pi \I\sum_A\frac{c_Aq_A}{N_A}\right)\,,
\qquad
\Big|[g]\Big\rangle\longmapsto\chi_c\Big(q(g)\Big)\Big|[g]\Big\rangle\,.
\label{eq:chargephase}
\end{equation}
\Cref{eq:abelian_selection} is equivalent to invariance under
all these characters. This implies that \Cref{eq:abelian_selection} must be amended as
\begin{equation}
\label{eq:AbelianizationConservationLaws}
  \sum_{a=1}^mq_A(g_a)=0\bmod{N_A}\qquad \text{for every }A\,.    
\end{equation}
The groups $\Abhat{S}$ and $\Ab(S)$ are abstractly isomorphic when finite,
but the former is the group whose elements are \U1 phases that act on string states.

To appreciate the charge conservation directly, consider a (superpotential) coupling among localized
fields $\mathcal I=\Phi_{[g_1]}\cdots\Phi_{[g_m]}$, suppressing its gauge
contractions. Under a phase transformation, it becomes
\begin{equation}
\mathcal I\longmapsto
\left[\prod_{a=1}^{m}\chi_c\Big(q(g_a)\Big)\right]\mathcal I
=\exp\!\left(2\pi \I\sum_A\frac{c_A}{N_A}
\sum_{a=1}^{m}q_A(g_a)\right)\mathcal I.
\label{eq:couplingchargephase}
\end{equation}
Every allowed coupling must obey the separate conservation laws in \Cref{eq:AbelianizationConservationLaws} for all independent charges. 

The relations among the generators
are essential: the exponents $n_i$ and $k_a$ in
\Cref{eq:chargefromword} are not generally independent conserved
charges. For instance, a relation $s(T_1)=s(T_2)$ makes a coupling sensitive
to $n_1+n_2$, while a relation $s(g_\omega)^2=s(T_1)$ can combine a
point-group exponent with a translation charge. We shall examine these possibilities with explicit examples in \Cref{sec:Abexamples}.

\subsection{The role of roto-translations in the charges}
\label{sec:systematic_abelianization}

Roto-translations affect the charges in two distinct ways. First, additional
commutators can close on lattice translations, forcing some lattice charges
to vanish. Second, relations such as \Cref{eq:liftpower} can tie a
surviving lattice charge to a point-group charge. The first effect reduces
the translation-charge subgroup, whereas the second can alter the cyclic
factorization of $\Ab(S)$ even when its order is unchanged.

The lattice charges removed by the first effect build the group
\begin{equation}
\operatorname{Im}d^2 := \frac{\Lambda\cap[S,S]}{\Lambda_P}
\quad\subseteq\quad\Lambda/\Lambda_P\,.
\label{eq:transgressionintersection}
\end{equation}
Here $\Lambda\cap[S,S]$ consists of lattice translations that can be written
as products of commutators in the full space group. The notation $d^2$ is standard for the map whose image, $\operatorname{Im}d^2$, identifies these additional relations in the low-degree homology associated with the short exact sequence~\eqref{eq:ext_space} defining the space group.\footnote{More
precisely, the Lyndon--Hochschild--Serre five-term exact sequence gives
$H_2(S,\ZZ)\to H_2(P,\ZZ)\xrightarrow{d^2}\Lambda/\Lambda_P\to
\Ab(S)\to\Ab(P)\to0$. Here $H_2(P,\ZZ)$ is the second group-homology group,
also called the Schur multiplier of $P$. For the five-term sequence, the
map $d^2$, and the relation with group extensions, see
Ref.~\cite{Brown:1982}.} 
After removing these trivial lattice charges, one obtains the short exact
sequence
\begin{equation}
0\longrightarrow
\frac{\Lambda/\Lambda_P}{\operatorname{Im}d^2}
\longrightarrow\Ab(S)\longrightarrow\Ab(P)\longrightarrow0\,.
\label{eq:short_exact}
\end{equation}
The left-hand group is the surviving translation-charge subgroup, while the
right-hand group is the point-group Abelianization. We thus see that 
the  order of the Abelian charge group is obtained by multiplying the lattice and point-group contributions and then dividing by the additional relations encoded in $\operatorname{Im}d^2$, i.e.\ 
$|\!\Ab(S)|=|\Lambda/\Lambda_P|\,|\Ab(P)|/|\operatorname{Im}d^2|$.

The sequence~\eqref{eq:short_exact} need not be a direct product. For example, if a nontrivial order-two translation charge satisfies
$s(g_\omega)^2=s(T_1)$ together with $s(T_1)^2=1$, then $s(g_\omega)$ has order four, as we show in 
\Cref{sec:Abexamples} (see e.g.\ $D_8\text{-}(1,2)$), 
where two order-two factors, which would be independent in the absence of roto-translations, are replaced by one $\Z4$ factor. 
Thus, the surviving lattice and point-group charges can be 
linked even after the trivial translation charges have been removed. Naturally, for a symmorphic space group, \Cref{eq:Absplit} is recovered. Note that a non-symmorphic space
group can nevertheless have the same abstract Abelian group as a symmorphic
one, due to the fact that the sequence~\eqref{eq:short_exact} may split, whether or not 
the sequence~\eqref{eq:ext_space} splits.\footnote{In practice, such a scenario can arise if a 
particular roto-translation changes the charge formula without modifying the group itself, see e.g.\ $S_3\text{-}(1,2)$ in \Cref{sec:Abexamples}.}

In practice, both effects can be read directly from the complete integer
relations in \Cref{eq:Abrelations}. Commutators of convenient pairs of
roto-translations can expose some of the extra relations, but not necessarily all of them.
We therefore use the full space-group presentation throughout. It is important to stress that no
fixed-locus data or affine normalizer is needed to determine $\Ab(S)$.

\subsection{Examples}
\label{sec:Abexamples}

\paragraph{The $S_3\text{-}(1,1)$ and $S_3\text{-}(1,2)$ geometries.}
For $S_3=\langle\omega,\theta\mid\omega^2=\theta^3=(\omega\theta)^2=1\rangle$,
the Abelianization in the symmorphic case is obtained from
\begin{equation}
s(g_\omega)^2=s(g_\theta)^3=s(g_\omega g_\theta)^2=1\,.
\label{eq:swt}
\end{equation}
We use multiplicative notation for $s$ in these examples. Since its images
commute, \Cref{eq:swt} implies $s(g_\theta)^2=1=s(g_\theta)^3$ and hence
$s(g_\theta)=1$. The surviving point-group charge is $k\bmod2$, i.e.\ $\Ab(P=S_3)=\Z2$.
For the first lattice class, the translation-conjugation relations
$g_aT_i g_a^{-1}=\prod_jT_j^{(\widehat\vartheta_a)_{ji}}$
in \Cref{eq:sgpresentation} give, after applying $s$,
\begin{equation}
\label{eq:relationsConjugationTi-S3}
\begin{aligned}
s(T_2)=s(T_1)^{-1}s(T_2)^{-1}\,,\qquad
&s(T_1)=s(T_2)\,,\\
s(T_3)^{-1}s(T_4)^2=1\,,\qquad
&s(T_3)s(T_4)=1\,,\\
s(T_5)^2=1\,,\qquad
&s(T_6)^2=1\,.
\end{aligned}
\end{equation}
For example, $g_\omega T_2g_\omega^{-1}=T_1^{-1}T_2^{-1}$
gives $s(T_2)=s(T_1)^{-1}s(T_2)^{-1}$, or equivalently
$s(T_1)s(T_2)^2=1$. Combining the first two relations yields
$s(T_2)^3=1$; the next two similarly give
$s(T_3)=s(T_4)^2$ and $s(T_4)^3=1$.
Hence, we can rearrange the relations in \Cref{eq:relationsConjugationTi-S3} as
\begin{equation}
\begin{gathered}
s(T_1)=s(T_2)\,,\qquad s(T_2)^3=1\,,\qquad
s(T_3)=s(T_4)^2\,,\qquad s(T_4)^3=1\,,\\
s(T_5)^2=s(T_6)^2=1\,.
\end{gathered}
\label{eq:S311translationcharges}
\end{equation}
Consequently, the two pairs $(T_1,T_2)$ and $(T_3,T_4)$
give two independent order-three charges:
\begin{equation}
s(T_1)^{n_1}s(T_2)^{n_2}=s(T_2)^{n_1+n_2}
\qquad\text{and}\qquad
s(T_3)^{n_3}s(T_4)^{n_4}=s(T_4)^{2n_3+n_4}\,,
\label{eq:S311chargeproducts}
\end{equation}
where the integers $n_i$ should be related to the values found in the constructing elements of states arising in the $S_3\text{-}(1,1)$ case.
The last two translations in \Cref{eq:S311translationcharges}
provide independent order-two charges $n_5$ and $n_6$. Thus, for a constructing element
$g=T_1^{n_1}\cdots T_6^{n_6}g_\omega^k g_\theta^\ell$, we obtain
\begin{equation}
\begin{aligned}
\Ab(S_3\text{-}(1,1))&\cong\Z2^3\x\Z3^2,\\
q&=(k,\,n_5,\,n_6;\ n_1+n_2,\,2n_3+n_4).
\end{aligned}
\label{eq:sys_s311}
\end{equation}
The first three entries are taken modulo two and the last two modulo three.
In particular, notice that there is no independent order-three charge $\ell$ from
$g_\theta$: it is removed by the mixed relation in \Cref{eq:swt}.

On the other hand, $S_3\text{-}(1,2)$ is generated by $g_\omega=(\omega,0)$ and
$g_\theta=(\theta,e_5/3)$ along with the translations $T_i$. The translation relations are unchanged, while
\begin{equation}
s(g_\omega)^2=1\,,\qquad s(g_\theta)^3=s(T_5)\,,\qquad
s(g_\omega g_\theta)^2=1\,.
\end{equation}
Since the images commute, the mixed relation again implies
$s(g_\theta)^2=1$. Combining this with $s(g_\theta)^3=s(T_5)$ gives
$s(g_\theta)=s(T_5)$, and hence
\begin{equation}
s(g_\theta)^\ell s(T_5)^{n_5}=s(T_5)^{\ell+n_5}.
\label{eq:S312chargeproduct}
\end{equation}
The roto-translation exponent $\ell$ and $n_5$ therefore enter the same
order-two charge. The other charges are unchanged, so that
\begin{equation}
\begin{aligned}
\Ab\big(S_3\text{-}(1,2)\big)&\cong\Z2^3\x\Z3^2,\\
q&=(k,\,\ell+n_5,\,n_6;\ n_1+n_2,\,2n_3+n_4).
\end{aligned}
\label{eq:sys_s312}
\end{equation}
The structure of the charge group is unchanged, but one charge now involves a 
roto-translation exponent. No translation charge is lost and no higher-order cyclic factor appears.\footnote{Here $H_2(S_3,\ZZ)=0$, so $d^2=0$.
Furthermore, the independent generators displayed above ensure that the sequence~\eqref{eq:short_exact} splits, 
although the space group itself is non-symmorphic.
}

\paragraph{The first $D_8$ lattice class.}
The order-eight dihedral group $D_8$ can be generated by two reflections satisfying the presentation
$D_8=\langle\omega,\theta\mid\omega^2=\theta^2=(\omega\theta)^4=1\rangle$. The common translation relations are
\begin{equation}
s(T_1)=s(T_3),\qquad s(T_2)=s(T_4)^{-1},\qquad
s(T_3)^2=s(T_4)^2=s(T_5)^2=s(T_6)^2=1\,.
\label{eq:D8translationcharges}
\end{equation}
Thus, we find $\Lambda/\Lambda_P\cong\Z2^4$ and $\Ab(D_8)\cong\Z2^2$.
With $g_\theta=(\theta,0)$, four choices of $g_\omega=(\omega,\mu_\omega)$
illustrate the two effects in \Cref{eq:short_exact}:
\begin{equation}
\begin{array}{c|c|ccc}
\text{Geometry}&\mu_\omega&g_\omega^2&g_\theta^2&(g_\omega g_\theta)^4\\\hline
D_8\text{-}(1,1)&0&1&1&1\\
D_8\text{-}(1,2)&e_1/2&T_1&1&1\\
D_8\text{-}(1,4)&e_5/4&1&1&T_5\\
D_8\text{-}(1,5)&e_1/2+e_5/4&T_1&1&T_5
\end{array}
\label{eq:D8liftrelations}
\end{equation}
In the symmorphic case $D_8\text{-}(1,1)$, all six charges are independent. In $D_8\text{-}(1,2)$,
$s(g_\omega)^2=s(T_1)$ instead produces an order-four generator. In
$D_8\text{-}(1,4)$, the fourth-power relation and the two involution relations
force $s(T_5)=1$. In $D_8\text{-}(1,5)$, both effects occur: the fourth-power
relation gives $s(T_5)=s(T_1)^2=1$, while $s(g_\omega)$ has order four.
The order-four charge is particularly transparent in either case where
$g_\omega^2=T_1$: using $s(T_1)=s(T_3)$, one obtains
\begin{equation}
s(g_\omega)^k s(T_1)^{n_1}s(T_3)^{n_3}
=s(g_\omega)^{k+2n_1+2n_3}.
\label{eq:D8chargeproduct}
\end{equation}
This explains both the coefficient two and the modulus four without
assigning an independent charge to each original generator.
The resulting groups and charges are
\begin{equation}
\begin{array}{c|c|l}
\text{Geometry}&\Ab(S)&\text{Charges}\\\hline
D_8\text{-}(1,1)&\Z2^6&k,\ell,n_1+n_3,n_2+n_4,n_5,n_6\\
D_8\text{-}(1,2)&\Z2^4\x\Z4&\ell,n_2+n_4,n_5,n_6;\ k+2n_1+2n_3\\
D_8\text{-}(1,4)&\Z2^5&k,\ell,n_1+n_3,n_2+n_4,n_6\\
D_8\text{-}(1,5)&\Z2^3\x\Z4&\ell,n_2+n_4,n_6;\ k+2n_1+2n_3
\end{array}
\label{eq:D8Abexamples}
\end{equation}
The charges preceding a semicolon are modulo two; the last entry after it
is modulo four. The four examples distinguish the effects of
roto-translations: $D_8\text{-}(1,2)$ combines two charges into an order-four
charge, $D_8\text{-}(1,4)$ removes the $n_5$ charge, and $D_8\text{-}(1,5)$ does both.
The first two groups have order $64$ and the last two have order
$32$. A coefficient two in an individual charge is not by itself evidence
that a translation charge has been removed; the complete set of relations
must be used. The charge assignments for all 331 geometries of non-Abelian orbifolds are given in Appendix~\ref{app:charges}.

\section{Geometric symmetries at generic moduli}
\label{sec:geometric}

\subsection{Affine normalizer in the two bases}

In addition to the symmetries arising from the Abelianization of the space group, there are some geometric transformations that can exchange localized states.
Such transformations must map every allowed string boundary
condition to another boundary condition of the same orbifold. For an affine
map $\alpha$, this means that conjugation maps the space group to itself;
the set of these maps is called its affine normalizer.
Throughout, the automorphism induced by
$\alpha$ is fixed to be
\begin{equation}
\phi(\alpha)\left(\overline{h}\right):=\overline{\alpha\,h\,\alpha^{-1}}\in\Ab(S)\,,
\label{eq:phidef}
\end{equation}
so that successive geometric transformations act in the same order on
the charges.\footnote{The convention $h\mapsto\alpha^{-1}h\alpha$
reverses the order of composition. It is useful for the normalizer test
below, but we keep \Cref{eq:phidef} for the group action.} Instead of $q(g)$, we find more useful here the notation $\overline{g}$ for the elements of $\Ab(S)$ described as the classes of elements of $g\in S$ identified by space-group commutation relations.
The corresponding action on the phase characters is
\begin{equation}
(\alpha\cdot\chi)\left(\overline{h}\right)
=
\chi\!\left(\phi(\alpha)^{-1}\overline{h}\right)
=\chi\Big(\overline{\alpha^{-1}h\,\alpha}\Big).
\label{eq:dualaction}
\end{equation}
Following Refs.~\cite{Baur:2020jwc,Baur:2019kwi}, we test membership in the affine normalizer in the
convenient form
\begin{equation}
\alpha^{-1}g\alpha\in S
\qquad\text{for all }g\in S .
\label{eq:normalizercoord}
\end{equation}
Thus, writing $\Aff(\RR^6)$ for the group of invertible affine maps of
the covering space, the affine normalizer is given by
\begin{equation}
N_{\Aff}(S)
:=
\{\alpha\in\Aff(\RR^6)\mid \alpha^{-1}S\alpha\stackrel!=S\}.
\label{eq:NAff}
\end{equation}
Note that we can equivalently require $\alpha S\alpha^{-1}\stackrel!=S$. In our work, the action on charges and localization labels is
$g\mapsto\alpha g\alpha^{-1}$, as in \Cref{eq:phidef}.
In the lattice basis, the normalizer condition is algebraic. So, for
$\widehat\alpha=(\widehat\sigma,t)$ and
$\widehat g=(\widehat\vartheta,\widehat\mu_\vartheta+n)$,
\begin{align}
\widehat\alpha\,\widehat g\,\widehat\alpha^{-1}
\!=\!
\left(
\widehat\vartheta',\,
\widehat\mu_{\vartheta'}+n'
\right)
\quad\text{with}\quad
\widehat\vartheta'
:=\widehat\sigma\widehat\vartheta\widehat\sigma^{-1}
\quad\text{\&}\quad
\widehat\mu_{\vartheta'}+n':=\widehat\sigma(\widehat\mu_\vartheta+n)
+(\Id-\widehat\vartheta')t.
\label{eq:affineconjugation}
\end{align}
Therefore, $\alpha\in N_{\Aff}(S)$ precisely when, for every
$\widehat\vartheta\in\widehat P$,
\begin{equation}
\begin{aligned}
&\widehat\sigma\in \GL{6,\ZZ},\qquad
\widehat\vartheta'
=
\widehat\sigma\widehat\vartheta\widehat\sigma^{-1}
\in\widehat P,\\
&
\widehat\sigma\widehat\mu_\vartheta
+(\Id-\widehat\vartheta')t
-\widehat\mu_{\vartheta'}=-\widehat{\sigma}n+n'
\in\ZZ^6 \,.
\end{aligned}
\label{eq:normalizercond}
\end{equation}
The first line preserves the lattice and normalizes the point group; the second is the affine-lift
condition for the full boundary conditions, including roto-translations. To retain only symmetries present at generic moduli we also impose\footnote{The search for all possible $\widehat\sigma$ is finite. The point-group-invariant,
positive-definite matrix
$\widehat G_*=\sum_{\widehat\vartheta\in\widehat P}
\widehat\vartheta^T\widehat\vartheta$
belongs to the space of allowed metrics. Hence, any
$\widehat\sigma$ satisfying \Cref{eq:univGB} also preserves
$\widehat G_*$ and, therefore, belongs to
$O(\widehat G_*,\ZZ)$. This group is finite because the
positive-definiteness of $\widehat G_*$ bounds the lengths
of the columns of $\widehat\sigma$, leaving only finitely
many possible integer matrices. We use this group $O(\widehat G_*,\ZZ)$ to enumerate the
candidate transformations.}
\begin{equation}
\widehat\sigma^T G_i\widehat\sigma=G_i\,,
\qquad
\widehat\sigma^T B_a\widehat\sigma=B_a
\qquad
\text{for all }i,a \,.
\label{eq:univGB}
\end{equation}
We denote the resulting group by $N_{\Aff}^{\rm geo}(S)$. Transformations
already in $S$ are orbifold identifications, so they do not give additional
geometric symmetries. Dividing them out results in
\begin{equation}
\Outgeo(S):=N_{\Aff}^{\rm geo}(S)/S \,.
\label{eq:OutgeoDef}
\end{equation}
This justifies the name {\it geometric outer symmetry/transformation} that we shall use here.\footnote{Conjugation by
an element of $S$ is an inner automorphism; outer automorphisms identify
actions that differ by one of these. Here the quotient in
\Cref{eq:OutgeoDef} has no additional kernel in its action by outer
automorphisms. An affine map commuting with all lattice translations has
linear part $\Id$; commuting also with every point-group lift requires
a common invariant translation vector, which vanishes in this catalogue.}

\subsection{Fractional translations \texorpdfstring{$T_P$}{TP}}
Setting $\widehat\sigma=\Id$ in \Cref{eq:normalizercond} gives
\begin{equation}
T_P=
\left\{
t\in\RR^6
\ \middle|\
(\Id-\widehat\vartheta)t\in\ZZ^6
\ \text{for all }\widehat\vartheta\in\widehat P
\right\}/\ZZ^6 .
\label{eq:TP}
\end{equation}
In the coordinate basis these transformations are simply
$\alpha_t=(\Id,e\,t)$.  They leave $G$ and $B$ untouched and depend only on the
linear point-group action.  Their invariant factors follow from the Smith normal form of the stacked
matrices $\Id-\widehat\vartheta$. The stack has rank six because the point-group
representation has no common invariant vector, so $T_P$ is finite.
Geometrically, $T_P$ consists of the fractional translations compatible with the
point-group action and can permute the inequivalent localization labels within
each twisted sector. Let $[p]$ run over the nontrivial point-group conjugacy
classes with localized sectors, and define 
\begin{equation}
\mathcal L(S):=\bigsqcup_{[p]}\mathcal L_{[p]} \,.
\end{equation}
Here, $\mathcal L_{[p]}$ is the set of inequivalent localization labels in the twisted sector $[p]$, and $\bigsqcup$ denotes their disjoint union over all localized twisted sectors. Thus, $\mathcal L(S)$ is simply the complete set of localization labels on which the geometric transformations act.

Interestingly, the localization labels can be grouped into geometric orbits. To see how this happens
it is useful to isolate the sector-preserving (sp) subgroup
\begin{equation}
\Outgeo^{\rm sp}(S):=
\left\{\alpha\in\Outgeo(S)\mid \alpha\Big([p]\Big)=[p]\text{ for every localized }[p]\right\}\,.
\label{eq:Outgeocp}
\end{equation}
These are the geometric outer transformations that preserve each twisted
sector separately.
Its natural maximal sector-by-sector permutation group is
\begin{equation}
S_{\rm naive}(\mathcal L)=\prod_{[p]}S_{|\mathcal L_{[p]}|}\,,
\end{equation}
since, before imposing the geometric constraints, the localization labels
within each sector could in principle be permuted independently.
The full geometric normalizer may, however, also
exchange distinct localized point-group conjugacy classes when a
geometric normalizer transformation maps one twisted sector to another. Accordingly,
\begin{equation}
\rho_{\mathcal L}(T_P)\triangleleft
\rho_{\mathcal L}(\Outgeo^{\rm sp})
\subseteq S_{\rm naive}(\mathcal L)\,,
\qquad
\rho_{\mathcal L}(\Outgeo)\subseteq S_{|\mathcal L|}\,,
\end{equation}
with possible kernels understood.

For our catalogue of 331 non-Abelian geometries, it is inconvenient to list
every set $\mathcal L_{[p]}$ explicitly. We therefore summarize the localization
data in two compact ways. If a sector representative has order $o_{[p]}$, its
connected fixed components have real dimension $d_{[p]}$, and
$N_{[p]}:=|\mathcal L_{[p]}|$, we can define the
\emph{sector localization profile}
\begin{equation}
\Sigma_{\mathcal L}(S)
:=\bigl\{\,o_{[p]}:N_{[p]}T^{d_{[p]}}\,\bigr\}_{[p]} .
\end{equation}
Repeated identical entries are compressed as $(o:NT^d)^{\x m}$.
This notation records, for each twisted sector, the twist order, the number
of inequivalent localization labels, and the dimension of the corresponding
fixed components, e.g.\ points ($T^0$) or fixed two-tori ($T^2$).

We also record the \emph{traditional geometric orbit profile}
$\mathcal O_{\rm geo}(S)$. If the
sector-preserving action of $\Outgeo^{\rm sp}(S)$
decomposes $\mathcal L(S)$ into $m_r$ orbits of size $r$, we use the shorthand
\begin{equation}
    \mathcal O_{\rm geo}(S)=1^{m_1}2^{m_2}3^{m_3}\cdots,
    \qquad
    |\mathcal L(S)|=\sum_{r\ge1}r\,m_r .
\end{equation}
Any exponent equal to one is omitted.
Thus, $\mathcal O_{\rm geo}(S)$ simply records how the localization labels
are grouped into orbits under the sector-preserving geometric symmetry.
Together, these two profiles provide complementary information:
$\Sigma_{\mathcal L}(S)$ summarizes the localization structure sector by
sector, while $\mathcal O_{\rm geo}(S)$ describes how the corresponding
localization labels are related by the sector-preserving geometric symmetry.
They therefore give a compact description of both the localized twisted
sectors and their geometric permutation structure. 

\subsection{Additional linear outer symmetries}

We write
$N_{\mathrm{GL}(6,\ZZ)}(\widehat P)=\{\widehat\sigma\in \mathrm{GL}(6,\ZZ)\mid
\widehat\sigma^{-1}\widehat P\widehat\sigma=\widehat P\}$ for the
ordinary linear normalizer of the point group. The background-preserving
part that admits an affine lift is
\begin{equation}
N_{\rm lin}^{\rm geo}(S)
:=
\left\{
\widehat\sigma\in N_{\mathrm{GL}(6,\ZZ)}(\widehat P)
\ \middle|\
\begin{array}{l}
\widehat\sigma^TG_i\widehat\sigma=G_i,\quad
\widehat\sigma^TB_a\widehat\sigma=B_a,\\[-1mm]
\exists\,t\ \text{such that }(\widehat\sigma,t)
\text{ satisfies \Cref{eq:normalizercond}}
\end{array}
\right\}.
\label{eq:Nlingeom}
\end{equation}
Point-group transformations are already part of the orbifold identification, so the remaining
linear factor is
\begin{equation}
\Outlin(S):=N_{\rm lin}^{\rm geo}(S)/\widehat P \,.
\label{eq:Outlin}
\end{equation}
The full geometric outer group is then organized as
\begin{equation}
1\longrightarrow T_P
\longrightarrow\Outgeo(S)
\longrightarrow\Outlin(S)
\longrightarrow1\,,
\qquad
|\Outgeo|=|T_P|\,|\Outlin|\,.
\label{eq:OutgeoSES}
\end{equation}
For a symmorphic space group every background-preserving linear normalizer element lifts with
$t=0$.  Roto-translations can remove some of these transformations because the second line of
\Cref{eq:normalizercond} may have no solution.
Together with the character action, the affine normalizer therefore gives
\begin{equation}
T_P\triangleleft\Outgeo(S)\,,\qquad
\Ggeo(S)=\Abhat{S}\rtimes_{\phi^\vee}\Outgeo(S)\,,
\label{eq:SnToOutgeo}
\end{equation}
with the full group acting on the union of all localized sectors, and $\phi^\vee$ 
is the induced action on the character group, $\phi^\vee(\alpha)(\chi)=\chi\circ\phi(\alpha)^{-1}$. 
Thus the symbol $\rtimes_{\phi^\vee}$ simply means that a geometric transformation
also transforms the phase character according to \Cref{eq:dualaction}.

\subsection{Where roto-translations enter}
Roto-translations do not change the allowed $G,B$ backgrounds and they do not change $T_P$.
They affect the result in two places.  First, they modify the relations of $S$ and hence can
change $\Ab(S)$.  Second, they can obstruct the affine lift of an otherwise allowed
$\widehat\sigma$, reducing $\Outlin$ and $\Outgeo$.  Both effects are obtained directly from
the space-group presentation and \Cref{eq:normalizercond}.

\section{Classical flavor groups and localization representations}
\label{sec:flavorgroups}
The construction naturally produces two nested groups.  The shift subgroup is
\begin{equation}
\label{eq:Gshift}
  \Gshift(S):=\Abhat{S}\rtimes_{\phi^\vee}T_P\,,
\end{equation}
while the maximal moduli-independent geometric parent is
\begin{equation}
\Ggeo(S):=\Abhat{S}\rtimes_{\phi^\vee}\Outgeo(S) \,.
\label{eq:Ggeo}
\end{equation}
Here $\phi^\vee$ is the induced action on the character group, defined as before; for $\Gshift$ the same action is restricted to $T_P$.
Hence,
\begin{equation}
\label{eq:ratio}
  \Gshift\subseteq\Ggeo\,,
  \qquad
  \frac{|\Ggeo|}{|\Gshift|}=|\Outlin| \,.
\end{equation}
The first group contains the phases and fractional translations; the second also includes
the allowed linear transformations.

For the localization-label set $\mathcal L$ defined above, we introduce the formal vector space $\mathcal V_{\mathcal L}$ with basis $\{e_{[g]}\}_{[g]\in\mathcal L}$. Here $e_{[g]}$ denotes the basis vector associated with the localization label $[g]$, rather than a lattice basis vector. Recalling that the image of $g$ in $\Ab(S)$ is denoted by $\overline g$, the two elementary actions are
\begin{equation}
\rho(\chi)e_{[g]}
:=\chi(\overline g)e_{[g]}\,,
\qquad
\rho(\alpha)e_{[g]}
:=e_{[\alpha g\alpha^{-1}]}\,.
\label{eq:classicalaction}
\end{equation}

Thus $\Abhat{S}$ is diagonal in a constructing-element basis, whereas
geometric outer automorphisms permute those basis labels. These two actions
satisfy
\begin{equation}
\label{eq:phasepermutationconjugation}
  \rho(\alpha)\rho(\chi)\rho(\alpha)^{-1}
  =\rho\!\left(\chi\circ\phi(\alpha)^{-1}\right)\,.
\end{equation}
Equation~\eqref{eq:phasepermutationconjugation} fixes the group structure,
not only its order, directly from the space-group relations and affine
conjugation. No twist-field operator products are required at this stage.
A gauge embedding, Wilson lines and the orbifold projection can reduce the
faithful symmetry on the physical spectrum; \Cref{eq:Gshift,eq:Ggeo} therefore define classical parent groups.

\subsection{The \texorpdfstring{$T^2/\Z2$}{T2/Z2} orbifold}
\label{sec:Z2example}
The $T^2/\Z2$ orbifold reproduces its known generic traditional
flavor group \cite{Baur:2020jwc}.  Let
\begin{equation}
\theta=-\Id_2\,,
\qquad
\widehat\theta=-\Id_2\,,
\qquad
S=\{(\theta^k,e\,n)\mid k=0,1,\ n\in\ZZ^2\}\,.
\end{equation}
The four twisted conjugacy classes can be labeled by
$n_1,n_2\in\{0,1\}$, with localized fields
\begin{equation}
\Phi=
\bigl(
\phi_{(0,0)},\phi_{(1,0)},\phi_{(0,1)},\phi_{(1,1)}
\bigr)^T \,.
\label{eq:Z2Phi}
\end{equation}
From \Cref{eq:TP}, we obtain in this case
\begin{equation}
2t\in\ZZ^2
\qquad\Longrightarrow\qquad
T_P\cong\Z2\x\Z2\,,
\end{equation}
generated by
\begin{equation}
\alpha_1=
\left(\Id_2,\tfrac12e_1\right)\,,
\qquad
\alpha_2=
\left(\Id_2,\tfrac12e_2\right)\,.
\end{equation}
On the localized fields they act as
\begin{equation}
\rho(\alpha_1)=
\begin{pmatrix}
0&1&0&0\\
1&0&0&0\\
0&0&0&1\\
0&0&1&0
\end{pmatrix},
\qquad
\rho(\alpha_2)=
\begin{pmatrix}
0&0&1&0\\
0&0&0&1\\
1&0&0&0\\
0&1&0&0
\end{pmatrix}.
\label{eq:Z2shifts}
\end{equation}
Thus, the two fractional translations exchange the fixed points pairwise.
The Abelianization follows immediately from
$\widehat\Lambda_P=2\ZZ^2$ and $\Ab(P)\cong\Z2$:
\begin{equation}
\Ab(S)
\cong
\Z2^2\x\Z2
=
\Z2^3 .
\label{eq:Z2Ab}
\end{equation}
A convenient set of character matrices is
\begin{equation}
Z_1=
\mathrm{diag}(1,-1,1,-1),
\qquad
Z_2=
\mathrm{diag}(1,1,-1,-1),
\qquad
Z_\theta=-\Id_4 .
\label{eq:Z2phases}
\end{equation}
The first two are the two translation-selection-rule phases and $Z_\theta$ is the point-group
phase.  Combining \Cref{eq:Z2shifts,eq:Z2phases} yields
\begin{equation}
\left\langle
\rho(\alpha_1),\rho(\alpha_2),Z_1,Z_2,Z_\theta
\right\rangle
\cong
\frac{D_8\x D_8}{\Z2^{\rm diag}}
\qquad\mathrm{ID}=[32,49]\,.
\label{eq:Z2fullgroup}
\end{equation}
Indeed, $\langle\rho(\alpha_1),Z_1\rangle\cong D_8$ and
$\langle\rho(\alpha_2),Z_2\rangle\cong D_8$, while their central elements are identified because
\begin{equation}
Z_\theta=
\bigl(\rho(\alpha_1)Z_1\bigr)^2
=
\bigl(\rho(\alpha_2)Z_2\bigr)^2 \,.
\end{equation}
At generic $T^2/\Z2$ moduli there is no further purely geometric linear factor after
quotienting by the point group, so
\begin{equation}
\Gshift=\Ggeo \cong
(D_8\x D_8)/\Z2 \,.
\end{equation}
Therefore, $T_P$ is the geometrically determined
$\Z2^2$ subgroup of the maximal sector-by-sector $S_4$ acting on the four fixed points.

\subsection{The \texorpdfstring{$T^2/\Z3$}{T2/Z3} orbifold}
For $T^2/\Z3$, we find
\begin{equation}
  \Abhat{S}\cong\Z3^2\,,
  \qquad
  T_P\cong\Z3\,,
  \qquad
  \Outlin\cong\Z2 \,.
\end{equation}
The fractional translation gives the three-cycle of the three fixed points, while the universal
$180^\circ$ rotation supplies the additional $\Z2$.  Therefore,
\begin{equation}
\label{eq:Z3benchmark}
  \Gshift\cong\Delta(27)
  \qquad\text{and}\qquad
  \Ggeo\cong\Delta(54)\,.
\end{equation}
The non-Abelian structure is already visible on the three localization
labels $e_r$, $r\in\Z3$. Choose the shift $X$ and a lattice-charge phase
$Z$ as
\begin{equation}
  X e_r=e_{r+1}\,,\qquad 
  Z e_r=\zeta^r e_r\,,\qquad
  ZX=\zeta XZ,\qquad \zeta=\e^{2\pi\I/3}\,.
\label{eq:Z3shiftphase}
\end{equation}
Their commutator gives the central phase $\zeta\mathbf1$, which represents
the point-group charge on this twisted sector. Thus $X$ and $Z$ generate
$\Delta(27)$, including both independent selection-rule phases. The
inversion $C e_r=e_{-r}$ gives $CXC^{-1}=X^{-1}$ and
$CZC^{-1}=Z^{-1}$, completing the group to $\Delta(54)$. This reproduces the familiar Narain result \cite{Baur:2019kwi,Baur:2019iai} directly from geometric permutations and Abelianization.

\subsection{The \texorpdfstring{$S_3\text{-}(3,1)$}{S3-(3,1)} orbifold}
\label{sec:S331example}
We consider $S_3\text{-}(3,1)$, the third integral lattice class and first affine
class of the six-dimensional $S_3$ point group. This symmorphic example
illustrates the distinction between localization labels and common fixed
points and admits a faithful representation of the complete parent.
In the lattice basis,
\begin{equation}
\widehat g_\omega=(\widehat\omega,0)\,,
\qquad
\widehat g_\theta=(\widehat\theta,0)\,,
\qquad
\widehat T_n=(\Id_6,n)\,,\quad n\in\ZZ^6\,,
\label{eq:S331generators}
\end{equation}
with
\begin{equation}
\widehat\omega^2=\widehat\theta^3
=(\widehat\omega\widehat\theta)^2=\Id_6 \,.
\label{eq:S331S3presentation}
\end{equation}
The corresponding coordinate-space generators are
\begin{equation}
 g_\omega=(\omega,0),\qquad
 g_\theta=(\theta,0),\qquad
 T_n=(\Id_6,e\,n),
\qquad
 \omega=e\,\widehat\omega e^{-1},\quad
 \theta=e\,\widehat\theta e^{-1}.
\label{eq:S331coordinate}
\end{equation}
For the crystallographic representative used in the scan of Refs.~\cite{Fischer:2012qj,Fischer:2013qza},
\begin{equation}
\widehat\omega=
\begin{pmatrix}
-1&0&0&0&0&0\\
0&0&-1&0&0&0\\
0&-1&0&0&0&0\\
0&0&0&0&-1&0\\
0&0&0&-1&0&0\\
0&0&0&0&0&-1
\end{pmatrix},
\qquad
\widehat\theta=
\begin{pmatrix}
1&0&0&0&0&0\\
0&-1&-1&0&0&0\\
0&1&0&0&0&0\\
0&0&0&0&1&0\\
0&0&0&0&0&-1\\
0&0&0&-1&0&0
\end{pmatrix}.
\label{eq:S331explicitgenerators}
\end{equation}
Here $\widehat e_i$ is the standard unit vector in the lattice basis and
$e_i=e\widehat e_i$ is the corresponding coordinate-space lattice vector.
\paragraph{Allowed $G,B$ backgrounds.}
The six-dimensional representation has
\begin{equation}
\chi(1)=6\,,\qquad
\chi(\text{order }2)=-2\,,\qquad
\chi(\text{order }3)=0\,.
\end{equation}
Hence, \Cref{eq:charcheck} yields
\begin{equation}
  n_G=6,\qquad n_B=2\,.
\label{eq:S331nGnB}
\end{equation}
Thus, a generic-moduli linear flavor transformation must preserve the complete six-dimensional
space of invariant metrics and the complete two-dimensional space of invariant $B$-fields.
\paragraph{Localization data and the common fixed set.}
For a point-group element $\widehat\vartheta$, the fixed locus on the torus is
\begin{equation}
\operatorname{Fix}_{T^6}(\vartheta)
=\left\{\widehat y\in\RR^6/\ZZ^6\ \middle|\
(\Id_6-\widehat\vartheta)\widehat y\in\ZZ^6\right\}\,.
\label{eq:S331sectorfixeq}
\end{equation}
This is a geometric subset of $T^6$.  A localized twisted-state label is instead associated with
a connected fixed component, equivalently with an appropriate space-group conjugacy class of a
constructing element.  These two notions coincide for isolated fixed points, but not when the fixed
locus contains a torus.
For the order-two representative,
\begin{equation}
\operatorname{SNF}(\Id_6-\widehat\omega)
=\operatorname{diag}(1,1,2,2,0,0),
\qquad
\ker(\Id_6-\widehat\omega)
=\operatorname{span}_{\RR}
\{\widehat e_3-\widehat e_2,\widehat e_5-\widehat e_4\}.
\label{eq:S331omegaSNF}
\end{equation}
Hence the $[\omega]$ sector contains four disconnected fixed two-tori,
\begin{equation}
\operatorname{Fix}_{T^6}(\omega)
=\bigcup_{a,b\in\{0,1\}}
\left[
\frac a2 e_1+\frac b2 e_6
+s(e_3-e_2)+t(e_5-e_4)
\right],
\qquad s,t\in\RR/\ZZ .
\label{eq:S331omegafixedtori}
\end{equation}
The two zero Smith entries give the two continuous directions; the nontrivial finite entries $2,2$ give the
four connected components.\footnote{The invariant direction $e_5-e_4$ should not be replaced by
$e_5+e_4$.  Although $\tfrac12(e_5-e_4)\equiv\tfrac12(e_5+e_4)\bmod\Lambda$, the two
parameterized curves differ by $2t\,e_4$, which is a lattice vector only at special values of
$t$.  Torus \emph{points} are identified modulo $\Lambda$; tangent directions are not vectors
modulo $\Lambda$.}
A convenient set of constructing elements for the four components is
\begin{equation}
\widehat g^{(\omega)}_{ab}
=(\widehat\omega,a\widehat e_1+b\widehat e_6),
\qquad a,b\in\Z2 .
\label{eq:S331omegaConstructing}
\end{equation}
For the order-three representative,
\begin{equation}
\operatorname{SNF}(\Id_6-\widehat\theta)
=\operatorname{diag}(1,1,1,3,0,0),
\qquad
\ker(\Id_6-\widehat\theta)
=\operatorname{span}_{\RR}
\{\widehat e_1,\widehat e_4+\widehat e_5-\widehat e_6\}.
\label{eq:S331thetaSNF}
\end{equation}
Thus, the $[\theta]$ sector contains three disconnected fixed two-tori,
\begin{equation}
\operatorname{Fix}_{T^6}(\theta)
=\bigcup_{r=0,1,2}
\left[
\frac r3(e_2+e_3)
+s e_1+t(e_4+e_5-e_6)
\right],
\qquad s,t\in\RR/\ZZ .
\label{eq:S331thetafixedtori}
\end{equation}
with convenient constructing elements
\begin{equation}
\widehat g^{(\theta)}_r=(\widehat\theta,r\widehat e_2),
\qquad r=0,1,2 .
\label{eq:S331thetaConstructing}
\end{equation}
The element $\theta^2$ belongs to the same nontrivial order-three conjugacy class of $S_3$, so
it does not define an additional point-group twisted sector.
At the classical localization level it is therefore natural to use the seven labels
\begin{equation}
\Psi_{\rm loc}
=\bigl(
|00\rangle_\omega,|10\rangle_\omega,|01\rangle_\omega,|11\rangle_\omega;
|0\rangle_\theta,|1\rangle_\theta,|2\rangle_\theta
\bigr)^T,
\label{eq:S331localizationbasis}
\end{equation}
where these labels refer to fixed-torus components/constructing-element classes, not to seven
isolated points.  A concrete heterotic gauge embedding can of course attach further gauge and
oscillator quantum numbers and can project some combinations.
The strict common fixed set is a different object:
\begin{equation}
\begin{aligned}
\mathcal F_{\rm com}\big(S_3\text{-}(3,1)\big)
&:=\operatorname{Fix}_{T^6}(\omega)\cap\operatorname{Fix}_{T^6}(\theta)\\
&=\left\{
0,\ \frac12e_1,\ \frac12(e_4+e_5+e_6),\
\frac12(e_1+e_4+e_5+e_6)
\right\},
\end{aligned}
\label{eq:S331commonpoints}
\end{equation}
so
\begin{equation}
n_{\rm com}=4.
\end{equation}
The common-point count differs from the number of localization labels: here $|\mathcal L_\omega|=4$, $|\mathcal L_\theta|=3$, while
$|\mathcal F_{\rm com}|=4$.
The incidence maps assign each common point to its connected fixed component.  In the present example,
\begin{equation}
\pi_\omega:\mathcal F_{\rm com}\xrightarrow{\sim}\mathcal L_\omega,
\qquad
\pi_\theta(\mathcal F_{\rm com})=
\bigl\{[\widehat g^{(\theta)}_0]\bigr\}
\subsetneq\mathcal L_\theta .
\label{eq:S331incidence}
\end{equation}
Thus, the four common points distinguish the four order-two fixed tori one by one, whereas all four
lie on the $r=0$ order-three fixed torus; the other two order-three localization components contain
no global common point.  The numerical equality $n_{\rm com}=4$ therefore has no interpretation
as a universal fourfold twisted-state degeneracy.
The construction acts on $\mathcal L$, the set of classical localization
labels. The common fixed set $\mathcal F_{\rm com}$ consists of target-space
positions, while the physical Hilbert space $\mathcal H_{\rm phys}$ also
requires the gauge embedding and orbifold projection.
\begin{equation}
\mathcal F_{\rm com}\longrightarrow\mathcal L_{[\vartheta]}
\quad\text{through incidence},\qquad
\mathcal L\quad\text{labels the basis of }\rho_{\mathcal L}.
\label{eq:S331threelevels}
\end{equation}

\paragraph{Sector-preserving permutations.}
Equations~\eqref{eq:S331incidence} and \eqref{eq:S331threelevels} also show why neither
$S_{n_{\rm com}}$ nor the undifferentiated $S_{|\mathcal L|}$ is an unrestricted physical permutation
group.  Here
\begin{equation}
\mathcal L=\mathcal L_\omega\sqcup\mathcal L_\theta,
\qquad
|\mathcal L_\omega|=4,
\qquad
|\mathcal L_\theta|=3,
\end{equation}
and the two sets belong to inequivalent order-two and order-three twisted sectors.  They therefore
cannot be arbitrarily mixed.  For this $S_3$ example the maximal sector-preserving permutation
group is
\begin{equation}
S_{\rm naive}(\mathcal L)
=S_{|\mathcal L_\omega|}\x S_{|\mathcal L_\theta|}
=S_4\x S_3
\subset S_7.
\label{eq:S331sectorwiseAmbient}
\end{equation}
For this example the two nontrivial classes have different orders, so the full geometric action is
automatically class preserving and satisfies
\begin{equation}
\rho_{\mathcal L}(T_P)\triangleleft
\rho_{\mathcal L}(\Outgeo)
\subseteq S_4\x S_3\,.
\label{eq:S331actualPermutation}
\end{equation}
In the compact notation used later in Table~\ref{tab:results331}, this geometry has
\[
\Sigma_{\mathcal L}=2:4T^2,\;3:3T^2,
\qquad
\mathcal O_{\rm geo}=1\,2\,4 .
\]
As shown below, $T_P\cong\Z2^2$ acts faithfully on the $[\omega]$ quartet and trivially on the
$[\theta]$ triplet. The remaining linear involution exchanges two labels of the triplet.

A useful contrast is the symmorphic geometry $S_3\text{-}(4,1)$ appearing in
Table~\ref{tab:results331}.  For its order-two and order-three representatives one finds
\begin{equation}
\operatorname{SNF}(\Id-\widehat\omega)
=\operatorname{diag}(1,1,2,2,0,0)\,,
\qquad
\operatorname{SNF}(\Id-\widehat\theta)
=\operatorname{diag}(1,1,3,3,0,0)\,.
\end{equation}
Hence,
\begin{equation}
\label{eq:S341localizationContrast}
  |\mathcal L_\omega|=4\,,
  \qquad
  |\mathcal L_\theta|=9\,,
  \qquad
  |\mathcal L|=13\,,
  \qquad
  S_{\rm naive}(\mathcal L)=S_4\x S_9\,,
\end{equation}
Accordingly, the compact catalogue entries are
\begin{equation}
\Sigma_{\mathcal L}\big(S_3\text{-}(4,1)\big)=2:4T^2,\;3:9T^2\,,
\qquad
\mathcal O_{\rm geo}\big(S_3\text{-}(4,1)\big)=4\,9 \,.
\end{equation}
The strict simultaneous-point count, by contrast, is
\begin{equation}
n_{\rm com}\big(S_3\text{-}(4,1)\big)=36\,.
\label{eq:S341ncomContrast}
\end{equation}
Thus even the largest value of $n_{\rm com}$ in the present scan is not a localization-state
multiplicity: here $36$ common positions coexist with only $4+9=13$ sector-wise localization
labels.  The common set should instead be viewed as incidence data for how the two families of fixed
two-tori intersect.

\paragraph{Abelian charges and phase representation.}
The surviving translation charges follow from
\begin{equation}
M_{\rm coin}
=\bigl(\Id_6-\widehat\omega\ \big|\
        \Id_6-\widehat\theta\bigr)\,,
\end{equation}
whose Smith invariants are
\begin{equation}
(1,1,1,1,2,6)\,.
\end{equation}
Therefore,
\begin{equation}
  \frac{\ZZ^6}{\widehat\Lambda_P}
  \cong\Z2\x\Z6
  \cong\Z2^2\x\Z3 \,.
\end{equation}
Since the space group is split and $\Ab(S_3)\cong\Z2$,
\begin{equation}
\label{eq:S331Ab}
  \Ab\big(S_3\text{-}(3,1)\big)\cong\Z2^3\x\Z3 \,.
\end{equation}
In the charge convention of Appendix~\ref{app:charges},
\begin{equation}
\begin{aligned}
q_k&=k\bmod2, &\qquad q_a&=n_1\bmod2\,,\\
q_b&=n_4+n_5+n_6\bmod2, &\qquad q_3&=2n_2+n_3\bmod3 \,.
\end{aligned}
\label{eq:S331charges}
\end{equation}
For a constructing element
$\widehat g=(\widehat\theta^{\ell}\widehat\omega^k,n)$, the corresponding character generators
act as
\begin{equation}
Z_k=(-1)^k,\quad
Z_a=(-1)^{n_1},\quad
Z_b=(-1)^{n_4+n_5+n_6},\quad
U_3=\zeta_3^{\,2n_2+n_3},
\quad \zeta_3=\e^{2\pi i/3}.
\label{eq:S331phaseaction}
\end{equation}
On the localization basis \eqref{eq:S331localizationbasis} this becomes the explicit block-diagonal
representation
\begin{equation}
\begin{aligned}
\rho(Z_k)&=(-\Id_4)\oplus\Id_3,\\
\rho(Z_a)&=\operatorname{diag}(1,-1,1,-1)\oplus\Id_3,\\
\rho(Z_b)&=\operatorname{diag}(1,1,-1,-1)\oplus\Id_3,\\
\rho(U_3)&=\Id_4\oplus\operatorname{diag}(1,\zeta_3^2,\zeta_3).
\end{aligned}
\label{eq:S3317Dphases}
\end{equation}
Thus the $[\omega]$ quartet carries the three $\Z2$ phases, while the independent $\Z3$
phase is visible on the $[\theta]$ triplet.

\paragraph{Fractional translations and the shift subgroup.}
The fractional translations are obtained from the vertically stacked matrix
\begin{equation}
M_T=\begin{pmatrix}
\Id_6-\widehat\omega\\
\Id_6-\widehat\theta
\end{pmatrix},
\end{equation}
whose nontrivial Smith invariants are $2,2$.  Hence
\begin{equation}
T_P\big(S_3\text{-}(3,1)\big)\cong\Z2\x\Z2\,.
\label{eq:S331TP}
\end{equation}
A choice of lattice-basis representatives is
\begin{equation}
t_1=\frac12\widehat e_1\,,
\qquad
t_2=\frac12(\widehat e_4+\widehat e_5+\widehat e_6)\,,
\qquad
\alpha_i=(\Id_6,e\,t_i)\,.
\label{eq:S331tgenerators}
\end{equation}
For a symmorphic orbifold there is a simple identification after an origin has been chosen:
\begin{equation}
\iota_0:T_P\longrightarrow T^6\,,
\qquad [t]\longmapsto [t]=\widehat\alpha_t(0)\,,
\label{eq:S331originmap}
\end{equation}
and its image is precisely the lattice-coordinate common set $\widehat{\mathcal F}_{\rm com}$ (cf.\ \Cref{eq:strictCommonFixed}).  Hence the four translation classes
$\{0,t_1,t_2,t_1+t_2\}$ are represented by the same four lattice-coordinate points as the strict
common set; coordinate-space points are obtained by multiplication with $e$.  This is an
identification of the underlying finite sets, not an identification of physical roles:
$\mathcal F_{\rm com}$ consists of positions, whereas $T_P$ consists of transformations.  In a
non-symmorphic affine class the former can be empty while the latter remains nontrivial.

Conjugation by a fractional translation gives
\begin{equation}
\widehat\alpha_t(\widehat p,n)\widehat\alpha_t^{-1}
=\bigl(\widehat p,n+(\Id-\widehat p)t\bigr)\,.
\label{eq:S331translationaction}
\end{equation}
On the order-two constructing elements, this yields
\begin{equation}
t_1:(a,b)\mapsto(a+1,b)\,,
\qquad
t_2:(a,b)\mapsto(a,b+1)\,,
\qquad a,b\in\Z2\,,
\label{eq:S331omegaShiftAction}
\end{equation}
whereas both translations act trivially on the three $[\theta]$ conjugacy classes.  For example,
$(\Id-\widehat\theta)t_2=\widehat e_5+\widehat e_6
=(\Id-\widehat\theta)\widehat e_6$, so the transformed constructing element remains in the
same space-group conjugacy class.

In the ordered quartet $(|00\rangle,|10\rangle,|01\rangle,|11\rangle)_\omega$,
\begin{equation}
X_1^{(4)}=
\begin{pmatrix}
0&1&0&0\\1&0&0&0\\0&0&0&1\\0&0&1&0
\end{pmatrix},
\qquad
X_2^{(4)}=
\begin{pmatrix}
0&0&1&0\\0&0&0&1\\1&0&0&0\\0&1&0&0
\end{pmatrix},
\end{equation}
so the full seven-dimensional localization representation is
\begin{equation}
\rho(t_1)=X_1^{(4)}\oplus\Id_3\,,
\qquad
\rho(t_2)=X_2^{(4)}\oplus\Id_3\,.
\label{eq:S3317Dshifts}
\end{equation}
The non-Abelianity of the shift subgroup is now visible directly.  The only nontrivial conjugation
relations needed are
\begin{equation}
\rho(t_1)\rho(Z_a)\rho(t_1)^{-1}=\rho(Z_kZ_a)\,,
\qquad
\rho(t_2)\rho(Z_b)\rho(t_2)^{-1}=\rho(Z_kZ_b)\,,
\label{eq:S331dualactionconcrete}
\end{equation}
while the crossed pairs commute and $U_3$ commutes with the two translations.  Thus the
$\Z2^3$ phases and $T_P\cong\Z2^2$ generate
\begin{equation}
\frac{D_8\x D_8}{\Z2^{\rm diag}}
\qquad\mathrm{ID}=[32,49]\,,
\label{eq:S331E32}
\end{equation}
and the independent $\Z3$ character gives
\begin{equation}
\Gshift\big(S_3\text{-}(3,1)\big)
\cong
\Z3\x\frac{D_8\x D_8}{\Z2^{\rm diag}}
\qquad\mathrm{ID}=[96,224]\,.
\label{eq:S331Gshift}
\end{equation}
\Cref{eq:S331dualactionconcrete} is the concrete realization, in this example, of the dual
action $\phi^\vee$ appearing in \Cref{eq:Gshift}.

\paragraph{The remaining linear involution.}
Imposing the complete generic-moduli conditions gives
\begin{equation}
|N_{\rm lin}^{\rm geo}(S)|=12\,,
\qquad
\Outlin\big(S_3\text{-}(3,1)\big)\cong\Z2\,,
\label{eq:S331Outlin}
\end{equation}
and therefore
\begin{equation}
1\longrightarrow\Z2^2
\longrightarrow\Outgeo\big(S_3\text{-}(3,1)\big)
\longrightarrow\Z2
\longrightarrow1\,,
\qquad |\Outgeo|=8\,.
\label{eq:S331Outgeo}
\end{equation}

The nontrivial linear coset is represented by
$\widehat R=-\Id_6$, with zero affine shift. It preserves every
$G$ and $B$, centralizes the point group, and represents an involution
outside $\widehat P$. It fixes the order-two localization classes and
sends $r\mapsto-r$ in the order-three sector. Therefore
\begin{equation}
  R U_3R^{-1}=U_3^{-1}\,,\qquad R^2=1\,,
\label{eq:S331Raction}
\end{equation}
while $R$ centralizes the order-32 shift and phase factor. Its localization
matrix is
\begin{equation}
\rho(R)=\Id_4\oplus
\begin{pmatrix}1&0&0\\0&0&1\\0&1&0\end{pmatrix}.
\label{eq:S3317DR}
\end{equation}
This also leads to $\Outgeo\big(S_3\text{-}(3,1)\big)\cong\Z2^3$.
Consequently, we see that
\begin{equation}
\Ggeo\big(S_3\text{-}(3,1)\big)
\cong
S_3\x\frac{D_8\x D_8}{\Z2^{\rm diag}}
\qquad\mathrm{ID}=[192,1524]\,.
\label{eq:S331Ggeo}
\end{equation}
The seven matrices in \Cref{eq:S3317Dphases,eq:S3317Dshifts,eq:S3317DR} therefore provide a faithful classical localization representation of the parent
group on $\mathbb C^4\oplus\mathbb C^3$.  This is a statement about the geometric localization
data, not a claim that the complete massless \ac{CFT} spectrum contains exactly seven fields.

\Cref{eq:S331Ab,eq:S331TP,eq:S331Gshift,eq:S331Ggeo} provide in the current case the charge, translation and flavor groups.
All seven localization labels enter the faithful representation, including
the order-three components that contain no common fixed point.

\paragraph{\CP and geometric $R$-symmetry transformations.}
Generic modular transformations are absent from the parent group in this case.
\CP-like transformations are likewise absent at generic moduli when they change the surviving
$B$-background, although they may reappear at special \CP-invariant loci.  A moduli-independent
linear transformation can also act nontrivially on the holomorphic three-form and therefore be of
geometric $R$-symmetry type.  For example, the representative $R=-\Id_6$ above acts as $-\Id_3$
on complex coordinates and reverses the holomorphic three-form. The catalogue
retains this geometric transformation; its lift to fermions must be specified
when extracting the strictly non-$R$ flavor group.

\section{Computation and results}
\label{sec:algorithm}

For each space group the computation can be organized as follows:
\begin{enumerate}
\item Read the affine generators in the lattice basis,
      $(\widehat\vartheta,\widehat\mu_\vartheta)$, and close their linear parts to obtain
      $\widehat P$.
\item Solve
      $\widehat\vartheta^TG\widehat\vartheta=G$ and
      $\widehat\vartheta^TB\widehat\vartheta=B$ to obtain the exact parameterizations in
      \Cref{eq:GBparam}, with $(n_G,n_B)$ given by \Cref{eq:charcheck}.
\item Build the positive-definite reference matrix
      $\widehat G_*=\sum_{\widehat\vartheta\in\widehat P}
      \widehat\vartheta^T\widehat\vartheta$, enumerate
      $O(\widehat G_*,\ZZ)$, and retain the $\widehat\sigma$ that normalize
      $\widehat P$ and preserve every $G_i,B_a$.
\item Compute $T_P$ from the Smith form of the stacked
      $\Id-\widehat\vartheta$ matrices.  For each surviving $\widehat\sigma$, solve the affine
      congruence in \Cref{eq:normalizercond}; this determines
      $N_{\rm lin}^{\rm geo}(S)$, $\Outlin(S)$, and $\Outgeo(S)$.
\item Compute $\Ab(S)$ from the complete space-group presentation and use
      \Cref{eq:dualaction} for its action on $\Abhat{S}$.
\item For each nontrivial localized point-group conjugacy class, compute the connected fixed
      components, quotient by the appropriate space-group/centralizer equivalence, and record
      $\Sigma_{\mathcal L}$ and the sector-preserving geometric orbit profile $\mathcal O_{\rm geo}$.
\item Construct $\Gshift$ and $\Ggeo$, export explicit generators to GAP, and identify the corresponding SmallGroup ID
      when available.
\end{enumerate}

All calculations are performed in the lattice basis.\footnote{When the finite
metric-preserving group is generated rather than listed explicitly, it must
be closed before the $B_a$ conditions are imposed; in particular
$-\Id_6$ must also be subjected to the affine-lift test.} Coordinate-basis
transformations are recovered by
$\vartheta=e\widehat\vartheta e^{-1}$ and
$\sigma=e\widehat\sigma e^{-1}$.

\subsection{Explicit representations of states due to their localization}
\label{sec:algorithm-rho}

The action of the parent groups on the localization labels is represented by monomial matrices, which
combine a permutation with charge phases.\footnote{A monomial matrix
has one nonzero entry in each row and each column; here every nonzero
entry is a root of unity. The image of the representation is the group
generated by these matrices. A faithful representation distinguishes
every parent-group element; its kernel consists of the elements
represented by the identity matrix.}
The representation is faithful when its image has the parent-group order; otherwise its kernel
records parent transformations that act trivially on the chosen localization labels.

For a nontrivial $\vartheta\in P$, set
$W=\Id-\widehat\vartheta$ and take a Smith decomposition $UWV=D$.  A constructing element
$(\widehat\vartheta,\tau)$ has a nonempty fixed locus if and only if $\tau\in\operatorname{im}_{\RR}W$, that is
\begin{equation}
\label{eq:locmembership}
(U\tau)_i=0 \quad\text{for every zero Smith entry } i \ (D_{ii}=0)\,,
\end{equation}
and since $(Un)_i$ runs over all of $\ZZ$ as $n$ runs over $\ZZ^6$, the twist class is
localized if and only if $(U\widehat\mu_\vartheta)_i\in\ZZ$ at every zero Smith entry. These entries correspond to unconstrained coordinates along the fixed locus; their number is
the dimension $d$ of the fixed torus recorded in $\Sigma_{\mathcal L}$.  Writing
$\tau=\widehat\mu_\vartheta+n$, the localization classes of the sector are labeled by the finite
coordinates
\begin{equation}
\mathrm{lab}(\widehat\vartheta,\tau)\!=\!\Big((U n)_i \bmod D_{ii}\Big)_{i:\,D_{ii}\neq0},
\quad
\widehat g_w=\Big(\widehat\vartheta,\ \widehat\mu_\vartheta+U^{-1}\widehat w\Big),
\quad
\widehat w_i=\begin{cases} w_i, & D_{ii}\neq0,\\[2pt] -(U\widehat\mu_\vartheta)_i, & D_{ii}=0.\end{cases}
\end{equation}
At the zero Smith entries, $\widehat w_i$ must cancel
$(U\widehat\mu_\vartheta)_i$. Setting these entries to zero is justified
only when the corresponding components of $U\widehat\mu_\vartheta$ vanish,
as they do for symmorphic lifts with zero shifts. Otherwise the resulting
constructing element has an empty fixed locus.  The label is unchanged by
conjugation with a pure lattice translation, because $UWm=DV^{-1}m$ vanishes modulo $D_{ii}$;
it is therefore enough to close the pairs $(\widehat\vartheta,w)$ under conjugation by the finitely many
$(\widehat p,\widehat\mu_p)$, $p\in P$.  The resulting equivalence classes are exactly the elements of
$\mathcal L$.  Acting with $\Outgeo^{\rm sp}$ gives the sizes recorded in
$\mathcal O_{\rm geo}$, while acting with the full $\Outgeo$ gives the permutation part of the
monomial representation below and is allowed to exchange point-group classes. To simplify notation, we use $\ell:=[g]$ as localization label. With this notation and using $q_\ell:=q(g_\ell)\in\Ab(S)$ the Abelianized charge of a chosen representative $g_\ell$ of $\ell\in\mathcal L$, the representation~\eqref{eq:classicalaction} takes the following form, suitable for explicit computation:
\begin{equation}
\rho(\chi)\,e_\ell=\chi(q_\ell)\,e_\ell\,,
\qquad
\rho(\alpha)\,e_\ell=e_{\alpha(\ell)}\,,
\qquad \chi\in\Abhat{S},
\qquad\alpha\in\Outgeo(S)\,,
\end{equation}
where $\alpha(\ell)$ denotes the label of $[\alpha g_\ell\alpha^{-1}]$, in the same direction as
\Cref{eq:phidef,eq:classicalaction}.  Representing a monomial matrix by the pair
$(\sigma,c)$, with $e_\ell\mapsto \e^{2\pi\I c_\ell}e_{\sigma(\ell)}$ and
$c_\ell\in\mathbb{Q}/\ZZ$ additive, the product is
$(\sigma_1,c_1)(\sigma_2,c_2)=(\sigma_1\!\circ\!\sigma_2,\ c_\ell=c_{2,\ell}+c_{1,\sigma_2(\ell)})$.
This representation allows exact finite-group closure. If only the order is needed, one can instead use the fact that
$\ker\rho$ consists of the pairs $(\chi,\alpha)$ with $\alpha$ acting trivially on
$\mathcal L$ and $\chi$ trivial on $Q$:
\begin{equation}
\label{eq:imrho}
\begin{aligned}
|\mathrm{Im}\,\rho|&=\frac{|Q|\;|\Outgeo(S)|}{|K_{\rm geo}|}\,,
&Q&=\langle q_\ell\rangle\subseteq\Ab(S)\,,\\
K_{\rm geo}&=\ker\big(\Outgeo(S)\to\mathrm{Sym}(\mathcal L)\big),
&K_T&=\ker\big(T_P\to\mathrm{Sym}(\mathcal L)\big)\,.
\end{aligned}
\end{equation}
where $\mathrm{Sym}(\mathcal L)$ denotes the permutation group of the
localization labels. The same formula for $\Gshift$ uses $K_T$ in place of
$K_{\rm geo}$. The two sources of a kernel can then be distinguished directly:
$Q\subsetneq\Ab(S)$ means that some nontrivial phase transformation acts
trivially on all localization labels, while $K_{\rm geo}\neq1$ means that
some nontrivial geometric transformation leaves every label fixed.

The localization condition removes translations along the invariant
directions of the twist, corresponding to nonzero winding along a fixed
torus. This is distinct from choosing a position on that torus: moving
within a connected fixed component does not change its constructing
element. For example, $e_3-e_2$ is fixed by $\omega$ in $S_3\text{-}(3,1)$ and
has a nontrivial Abelianized translation charge. Adding it to the
translation part of an $\omega$ constructing element introduces invariant
winding and makes \Cref{eq:locmembership} fail. The image groups in
Appendix~\ref{app:imagegroups} concern the localized boundary conditions
defined above. These groups correspond to the maximal faithful moduli-independent flavor groups that are compatible with the available massless states in the orbifold.

\subsection{Results for the 331 geometries}
\label{sec:results}
We apply this construction to the 331 affine geometries of
Refs.~\cite{Fischer:2012qj,Fischer:2013qza}. The 35 abstract point groups
and their Abelianizations are listed in Appendix~\ref{app:pointgroups},
and all space-group charge assignments appear in Appendix~\ref{app:charges}.
Table~\ref{tab:results331} collects the parent groups of moduli-independent flavor symmetries along with their (geometric) components, which include the symmetries arising from fractional translations, unbroken linear outer transformations, localization signatures, and sector-preserving orbit profiles.
The faithful moduli-independent flavor groups are listed separately in
Appendix~\ref{app:imagegroups}.

The charge and geometric columns are computed independently: $\Ab(S)$
follows from the affine presentation, $T_P$ from the stacked
$(\Id-\widehat\vartheta_a)$ congruences, and the remaining columns from
the affine normalizer and fixed loci. For non-symmorphic classes, the
roto-translations must enter both the presentation relations and the
fixed-locus congruences.
For every geometry the parent orders are
\begin{equation}
|\Gshift|=|\Ab(S)|\,|T_P|
\qquad\text{and}\qquad
|\Ggeo|=|\Ab(S)|\,|T_P|\,|\Outlin|\,,
\label{eq:catalogueorders}
\end{equation}
while the localization signature and orbit profile obey
\begin{equation}
\sum_{[p]}|\mathcal L_{[p]}|=\sum_{r\geq1}r\,m_r\,.
\label{eq:catalogueorbits}
\end{equation}
The actual semidirect-product structure is fixed by the action on the charge
characters, not by these orders alone.

Of the 662 parent entries, 518 have GAP SmallGroup identifiers. The remaining
144 entries are written $[|G|,?]$: 73 at order $512$, 45 at $1024$,
18 at $2048$, one at $3072$, three at $4096$, one at $6144$, two at
$11664$ and one at $24576$. Here a question mark records an unavailable
identification in the computation used for the catalogue; it does not
assert that the group has no catalogue identifier. In particular, the
availability of a group in the GAP SmallGroups library and the availability
of the automatic identification algorithm are separate questions
\cite{GAP4, SmallGrpManual}. No identifier is assigned from the group order alone.
\paragraph{Geometric factors and localization orbits.}
The localization signature $\Sigma_{\mathcal L}$ records the connected
components by twisted sector. The orbit profile $\mathcal O_{\rm geo}$
records their orbits under the sector-preserving geometric subgroup.
The full $\Outgeo$ exchanges
distinct point-group classes in 90 of the 331 geometries; those exchanges are included in the parent
and image groups but, by definition, not in $\mathcal O_{\rm geo}$.  Values $|\Outlin|>1$ signal additional linear
symmetries beyond the shift subgroup.  The pair $D_8\text{-}(1,1)$ and $D_8\text{-}(1,4)$, for example,
illustrates how a roto-translation can leave $T_P$ unchanged while modifying both $\Ab(S)$ and
the set of liftable linear transformations.  
A specific gauge embedding can reduce the symmetry visible on physical states.

As an illustrative non-symmorphic example, $D_8\text{-}(9,5)$ (the crystallographic
representative labeled \texttt{4682\_9\_5} in the CARAT-based database of
Refs.~\cite{Fischer:2012qj,Fischer:2013qza}) has four localized point-group
conjugacy classes with
\begin{equation}
\Sigma_{\mathcal L}
=(2:4T^2)^{\x2};\,2:5T^2;\,4:2T^2,
\qquad
|\mathcal L|=4+4+5+2=15,
\label{eq:D895audit}
\end{equation}
while the sector-preserving geometric action gives
\begin{equation}
\mathcal O_{\rm geo}=1^3\,2^2\,4^2,
\qquad
3+2\cdot2+2\cdot4=15.
\end{equation}
The order-four sector contributes two of these 15 labels. By contrast,
the strict common fixed set is empty, $n_{\rm com}=0$
(Appendix~\ref{app:commonfixed}).

\section{Conclusions and outlook}
\label{sec:conclusions}

We have developed a uniform geometric construction of the traditional or moduli-independent flavor symmetries for arbitrary space groups.
The method identifies two pieces
that are often treated independently: the space-group Abelianization
$\Ab(S)$, which determines the additive selection-rule charges, and the affine
transformations that preserve the full space group and every allowed
$G,B$ background, required to determine the geometric permutations that survive at generic moduli values. Their action on the phase group $\Abhat{S}$ then fixes both parent groups 
of the moduli-independent flavor group, $\Gshift=\Abhat{S}\rtimes_{\phi^\vee}T_P$ and
$\Ggeo=\Abhat{S}\rtimes_{\phi^\vee}\Outgeo(S)$. We observe that $\Ggeo$ includes modular flavor symmetries that leave the moduli invariant. By identifying the faithful representations of the localized states in the orbifold under the generators of the parent groups, we have finally determined the largest moduli-invariant flavor groups compatible with massless twisted string states of the orbifold.

A main structural result is the treatment of roto-translations, which had not been properly addressed before. They can
remove lattice charges through additional space-group relations, combine the
surviving lattice and point-group charges into larger cyclic factors, and
prevent an otherwise allowed linear transformation from lifting to a
symmetry of the full affine space group. This explains why naive approaches that rely on the fixed-component multiplicities, 
or results for symmorphic space groups cannot capture the richness found in non-Abelian orbifolds. 
The same construction treats point-group conjugacy classes, mixed generator relations and fixed 
tori without introducing separate case-by-case rules. In this sense, it extends the familiar 
phase-and-permutation picture of Abelian orbifolds to the non-Abelian classification while 
remaining entirely at the level of the space group and geometry.

For each of the 331 geometries we determined the Abelian charge group and explicit
charge representatives, the fractional translations $T_P$, the remaining
moduli-independent linear factor, the localization data by sector, the
geometric orbit structure, which build the two moduli-independent flavor parent groups. We also computed
their monomial actions on the localized constructing-element classes, so the
abstract parent can be distinguished from the group that acts faithfully on
these classical labels. This distinction is substantial: $\Gshift$ acts
faithfully in 254 geometries and $\Ggeo$ in 130, while the full geometric
outer group exchanges distinct point-group conjugacy classes in 90 of the
331 cases. The resulting catalogue therefore records not only the group structures, but
also the charges, permutations and localization data needed
to use those groups in explicit constructions. The final outcome of this exploration is the list of admissible moduli-independent flavor symmetry groups, presented in \Cref{app:imagegroups}. The specific charges are given in detail in other Appendices.

From a bottom-up perspective, our catalogue provides a UV-motivated guide to
discrete flavor model building. Phase symmetries, family permutations and
their mutual action are no longer independent {\it ad hoc} assumptions, but correlated
consequences of a common compactification geometry. From the string
phenomenology side, the 331 geometries form a detailed flavor landscape of
non-Abelian orbifolds and provide a practical starting point for model
building. Once a gauge embedding and Wilson lines are specified, the data
collected here can be combined with the orbifold projection and the remaining
string selection rules to determine the flavor representations of physical
massless states and, subsequently, the allowed Yukawa couplings and their
moduli dependence. The present work therefore supplies the geometric layer
of a systematic route from the compactification data to the flavor structure
of the four-dimensional effective field theory.

The construction also clarifies how this geometric picture sits inside the
Narain description. The fractional translations $T_P$ reproduce the
geometric part of the Narain translation sector, while characters of the
surviving lattice charges reproduce the corresponding dual phases; the full
space-group relations add the compatible point-group phases.
A more explicit comparison with the Narain-space-group normalizer is
given in \Cref{app:Narain}. A complete CFT
treatment is still required once one asks how the relevant group acts on a 
physical spectrum. Natural extensions are therefore to incorporate explicit
gauge embeddings and Wilson lines, separate ordinary flavor from geometric
$R$ symmetries, study modular and \CP enhancements at special loci, and extend
the analysis to asymmetric orbifolds. Similarly, one may ask the question 
whether our method can be extended to the study of modular flavor symmetries, 
which are known to be relevant in heterotic orbifolds~\cite{Lauer:1990tm,Lauer:1989ax,Baur:2024qzo}. 
Our catalogue may also provide a concrete string setting where one may ask broader quantum-gravity questions
about discrete symmetries, such as the absence of exact global symmetries and
charge completeness~\cite{Banks:2010zn,Harlow:2018jwu}. Such questions depend
on the full spectrum and on the gauge realization of the discrete symmetry,
and are therefore beyond the geometric parent groups determined here. 

\paragraph{Acknowledgments.}
This work was partly supported by UNAM-PAPIIT IN117226. The work of X-G.L. is supported by
Universidad Nacional Aut\'onoma de M\'exico Postdoctoral Program (POSDOC).
While the ideas, methods and algorithms presented in this work were conceived and developed by the authors, we acknowledge the use of Claude and ChatGPT to assist with routine tasks and some calculations, all of which were independently verified by the authors.
This work is dedicated to the loving memory of our dearest friend and collaborator Omar P\'erez-Figueroa, 
whose insight, kindness and enthusiasm will always shine brightly in our lives.

\appendix
\crefalias{section}{appendix}
\section{Abelianization and discrete charges}
\label{app:abelian_catalogue}

This appendix collects the point-group and space-group Abelianizations
and the discrete charges used in \Cref{sec:abelianization}.

\subsection{Point-group Abelianizations}
\label{app:pointgroups}
Table~\ref{tab:pointgroups} lists the 35 non-Abelian point groups in the
crystallographic classification, with $D_n$ denoting a dihedral group of
order $n$. Here $[P,P]$ is the derived subgroup generated by commutators and
$\Ab(P)=P/[P,P]$. Bracketed pairs specify GAP SmallGroup~\cite{SmallGrpManual} identifiers, where the first number is the order of the finite group and the second a counter of groups of such order; the
derived-subgroup and quotient identifiers are obtained by GAP~\cite{GAP4}.
\begingroup\small
\renewcommand{\arraystretch}{.99}
\setlength{\tabcolsep}{3pt}
\begin{longtable}{>{\raggedright\arraybackslash}p{.25\textwidth}l>{\raggedright\arraybackslash}p{.18\textwidth}l>{\raggedright\arraybackslash}p{.12\textwidth}l}
\caption{Point groups, derived subgroups and Abelianizations. Each bracketed pair in the ``ID'' columns is a GAP SmallGroup identifier.}\label{tab:pointgroups}\\
\toprule
$P$ & ID & $[P,P]$ & ID & $\Ab(P)$ & ID\\\midrule
\endfirsthead
\toprule $P$ & ID & $[P,P]$ & ID & $\Ab(P)$ & ID\\\midrule
\endhead\bottomrule\endfoot
$S_{3}$ & [6,1] & $\Z3$ & [3,1] & $\Z2$ & [2,1] \\
$D_{8}$ & [8,3] & $\Z2$ & [2,1] & $\Z2\x \Z2$ & [4,2] \\
$A_{4}$ & [12,3] & $\Z2\x \Z2$ & [4,2] & $\Z3$ & [3,1] \\
$D_{12}$ & [12,4] & $\Z3$ & [3,1] & $\Z2\x \Z2$ & [4,2] \\
$\Z8\rtimes \Z2$ & [16,6] & $\Z2$ & [2,1] & $\Z4\x \Z2$ & [8,2] \\
$QD_{16}$ & [16,8] & $\Z4$ & [4,1] & $\Z2\x \Z2$ & [4,2] \\
$(\Z4\x \Z2)\rtimes \Z2$ & [16,13] & $\Z2$ & [2,1] & $\Z2\x \Z2\x \Z2$ & [8,5] \\
$\Z3\x S_{3}$ & [18,3] & $\Z3$ & [3,1] & $\Z6$ & [6,2] \\
$T_{7}$ & [21,1] & $\Z7$ & [7,1] & $\Z3$ & [3,1] \\
$\Z3\rtimes \Z8$ & [24,1] & $\Z3$ & [3,1] & $\Z8$ & [8,1] \\
$\SL{2,3}\text{-I}$ & [24,3] & $Q_{8}$ & [8,4] & $\Z3$ & [3,1] \\
$\Z4\x S_{3}$ & [24,5] & $\Z3$ & [3,1] & $\Z4\x \Z2$ & [8,2] \\
$(\Z6\x \Z2)\rtimes \Z2$ & [24,8] & $\Z6$ & [6,2] & $\Z2\x \Z2$ & [4,2] \\
$\Z3\x D_{8}$ & [24,10] & $\Z2$ & [2,1] & $\Z6\x \Z2$ & [12,5] \\
$\Z3\x Q_{8}$ & [24,11] & $\Z2$ & [2,1] & $\Z6\x \Z2$ & [12,5] \\
$S_{4}$ & [24,12] & $A_{4}$ & [12,3] & $\Z2$ & [2,1] \\
$\Delta (27)$ & [27,3] & $\Z3$ & [3,1] & $\Z3\x \Z3$ & [9,2] \\
$(\Z4\x \Z4)\rtimes \Z2$ & [32,11] & $\Z4$ & [4,1] & $\Z4\x \Z2$ & [8,2] \\
$\Z3\x (\Z3\rtimes \Z4)$ & [36,6] & $\Z3$ & [3,1] & $\Z{12}$ & [12,2] \\
$\Z3\x A_{4}$ & [36,11] & $\Z2\x \Z2$ & [4,2] & $\Z3\x \Z3$ & [9,2] \\
$\Z6\x S_{3}$ & [36,12] & $\Z3$ & [3,1] & $\Z6\x \Z2$ & [12,5] \\
$\Delta (48)$ & [48,3] & $\Z4\x \Z4$ & [16,2] & $\Z3$ & [3,1] \\
$\GL{2,3}$ & [48,29] & $\SL{2,3}$ & [24,3] & $\Z2$ & [2,1] \\
$\SL{2,3}\rtimes \Z2$ & [48,33] & $Q_{8}$ & [8,4] & $\Z6$ & [6,2] \\
$\Delta (54)$ & [54,8] & $\Delta(27)$ & [27,3] & $\Z2$ & [2,1] \\
$\Z3\x \SL{2,3}$ & [72,25] & $Q_{8}$ & [8,4] & $\Z3\x \Z3$ & [9,2] \\
$\Z3\x((\Z6\x\Z2)\rtimes\Z2)$ & [72,30] & $\Z6$ & [6,2] & $\Z6\x \Z2$ & [12,5] \\
$\Z3\x S_{4}$ & [72,42] & $A_{4}$ & [12,3] & $\Z6$ & [6,2] \\
$\Delta (96)$ & [96,64] & $\Delta(48)$ & [48,3] & $\Z2$ & [2,1] \\
$\SL{2,3}\rtimes \Z4$ & [96,67] & $\SL{2,3}$ & [24,3] & $\Z4$ & [4,1] \\
$\Sigma (36\phi )$ & [108,15] & $\Delta(27)$ & [27,3] & $\Z4$ & [4,1] \\
$\Delta (108)$ & [108,22] & $\Z6\x \Z2$ & [12,5] & $\Z3\x \Z3$ & [9,2] \\
$\mathrm{PSL}(3,2)$ & [168,42] & $\mathrm{PSL}(3,2)$ & [168,42] & $1$ & [1,1] \\
$\Sigma (72\phi )$ & [216,88] & $\Delta(54)$ & [54,8] & $\Z2\x \Z2$ & [4,2] \\
$\Delta (216)$ & [216,95] & $\Delta(108)$ & [108,22] & $\Z2$ & [2,1] \\
\end{longtable}
\endgroup

\subsection{Space groups and discrete charges}
\label{app:charges}
\Cref{tab:charges331} lists the Abelianization and independent charges
for all 331 affine geometries of non-Abelian orbifolds compatible with heterotic compactifications with $\mathcal N=1$ supersymmetry in four dimensions. A geometry is denoted by
$P\text{-}(i,j)$, where $i$ labels the integral lattice class and $j$ its
affine class. The lattice basis and the ordered affine generators are those
of the crystallographic representatives used in
Refs.~\cite{Fischer:2012qj,Fischer:2013qza,Hernandez-Segura:2025sfr}.
The integers $k,\ell,m$ denote the exponents of the first, second and,
when needed, third chosen point-group lifts; $n_i$ are the coefficients
of the translation placed to their left. More general words are evaluated
using \Cref{eq:Abrelations}. Thus, the displayed formulae are basis-dependent charge representatives, while $\Ab(S)$ is basis independent.
Charges are listed in the order of the cyclic factors, with repeated
factors expanded. Each charge is reduced modulo the corresponding cyclic
order. Independent changes of character basis, including multiplication
by a unit modulo that order, give equivalent assignments. In particular,
the $D_8\text{-}(9,1)$ entry is written with the integral order-four charge
$2n_2+2n_3+2n_4+n_5+3n_6$.
\begingroup\footnotesize
\setlength{\tabcolsep}{3pt}
\renewcommand{\arraystretch}{1.12}
\begin{longtable}{>{\raggedright\arraybackslash}p{.25\textwidth}
>{\centering\arraybackslash}p{.18\textwidth}
>{\raggedright\arraybackslash}p{.51\textwidth}}
\caption{Space-group Abelianizations and independent discrete charges for all 331 geometries. Charges follow the displayed order of cyclic factors.}\label{tab:charges331}\\
\toprule Geometry & $\Ab(S)$ & Charges\\\midrule
\endfirsthead
\toprule Geometry & $\Ab(S)$ & Charges\\\midrule
\endhead\bottomrule\endfoot
$S_3\text{-}(1,1)$ & $\Z2^3\x \Z3^2$ & $k$, $n_{5}$, $n_{6}$, $n_{1}+n_{2}$, $2n_{3}+n_{4}$ \\
$S_3\text{-}(1,2)$ & $\Z2^3\x \Z3^2$ & $k$, $\ell +n_{5}$, $n_{6}$, $n_{1}+n_{2}$, $2n_{3}+n_{4}$ \\
$S_3\text{-}(2,1)$ & $\Z2^3\x \Z3$ & $k$, $n_{5}$, $n_{6}$, $2n_{1}+n_{2}$ \\
$S_3\text{-}(2,2)$ & $\Z2^3\x \Z3$ & $k$, $\ell +n_{5}$, $n_{6}$, $2n_{1}+n_{2}$ \\
$S_3\text{-}(3,1)$ & $\Z2^3\x \Z3$ & $k$, $n_{1}$, $n_{4}+n_{5}+n_{6}$, $2n_{2}+n_{3}$ \\
$S_3\text{-}(3,2)$ & $\Z2^3\x \Z3$ & $k$, $\ell +n_{1}$, $n_{4}+n_{5}+n_{6}$, $2n_{2}+n_{3}$ \\
$S_3\text{-}(4,1)$ & $\Z2^3$ & $k$, $n_{5}$, $n_{6}$ \\
$S_3\text{-}(4,2)$ & $\Z2^3$ & $k$, $\ell +n_{5}$, $n_{6}$ \\
\end{longtable}
\endgroup

\begin{landscape}

\section{Complete geometric flavor catalogue}
\label{app:fullcatalogue}

The following table collects the charge groups, geometric outer factors,
localization data and classical parents defined in the main text.
The $\Ab(S)$ column is written in exactly the same cyclic decomposition
as Table~\ref{tab:charges331}; the corresponding charge formulas are given
in Appendix~\ref{app:charges}.

\begingroup
\small
\setlength{\tabcolsep}{0.8pt}
\setlength{\LTleft}{0pt}
\setlength{\LTright}{0pt}
\setlength{\LTcapwidth}{1.22\textwidth}
\renewcommand{\arraystretch}{1.02}
\begin{longtable}{
>{\raggedright\arraybackslash}p{0.285\textwidth}
>{\centering\arraybackslash}p{0.09\textwidth}
>{\centering\arraybackslash}p{0.09\textwidth}
>{\centering\arraybackslash}p{0.095\textwidth}
>{\centering\arraybackslash}p{0.3\textwidth}
>{\centering\arraybackslash}p{0.1\textwidth}
>{\centering\arraybackslash}p{0.12\textwidth}
>{\centering\arraybackslash}p{0.12\textwidth}}
\caption{Classical geometric flavor data for all 331 non-Abelian affine geometries.
The labels are the same as in Table~\ref{tab:charges331}. In
$\Sigma_{\mathcal L}$, $o:NT^d$ denotes a twist class of order $o$ with
$N$ localization labels supported on $d$-dimensional fixed components;
$\x m$ compresses $m$ identical entries. The profile
$\mathcal O_{\rm geo}=1^{m_1}2^{m_2}\cdots$ gives the sector-preserving
geometric orbits. $\Outlin$ is the remaining linear outer group.
The final columns give the parent-group SmallGroup GAP identifiers;
$[|G|,?]$ denotes an order without an identified catalogue number.}
\label{tab:results331}\\
\toprule
Geometry & $\Ab(S)$ & $T_P$ & $\Outlin$ & $\Sigma_{\mathcal L}$ & $\mathcal O_{\rm geo}$ &  $\Gshift$ &  $\Ggeo$\\
\midrule
\endfirsthead
\toprule
Geometry & $\Ab(S)$ & $T_P$ & $\Outlin$ & $\Sigma_{\mathcal L}$ & $\mathcal O_{\rm geo}$ & $\Gshift$ & $\Ggeo$\\
\midrule
\endhead
\bottomrule
\endfoot
$S_3\text{-}(1,1)$ & $\Z2^3\x \Z3^2$ & $\Z2^2$ & $\Z2^{2}$ & $2:4T^{2};\,3:9T^{2}$ & $1\,2^{4}\,4$ & [288,\,1022] & [1152,\,153220] \\
$S_3\text{-}(1,2)$ & $\Z2^3\x \Z3^2$ & $\Z2^2$ & $\Z2$ & $2:4T^{2}$ & $4$ & [288,\,1022] & [576,\,8639] \\
$S_3\text{-}(2,1)$ & $\Z2^3\x \Z3$ & $\Z2^2\!\x\!\Z3$ & $\Z2^{2}$ & $2:4T^{2};\,3:9T^{2}$ & $3\,4\,6$ & [288,\,1022] & [1152,\,153220] \\
$S_3\text{-}(2,2)$ & $\Z2^3\x \Z3$ & $\Z2^2\!\x\!\Z3$ & $\Z2$ & $2:4T^{2}$ & $4$ & [288,\,1022] & [576,\,8639] \\
$S_3\text{-}(3,1)$ & $\Z2^3\x \Z3$ & $\Z2^2$ & $\Z2$ & $2:4T^{2};\,3:3T^{2}$ & $1\,2\,4$ & [96,\,224] & [192,\,1524] \\
$S_3\text{-}(3,2)$ & $\Z2^3\x \Z3$ & $\Z2^2$ & $1$ & $2:4T^{2}$ & $4$ & [96,\,224] & [96,\,224] \\
$S_3\text{-}(4,1)$ & $\Z2^3$ & $\Z2^2\!\x\!\Z3^{2}$ & $\Z2^{2}$ & $2:4T^{2};\,3:9T^{2}$ & $4\,9$ & [288,\,1022] & [1152,\,153220] \\
$S_3\text{-}(4,2)$ & $\Z2^3$ & $\Z2^2\!\x\!\Z3^{2}$ & $\Z2$ & $2:4T^{2}$ & $4$ & [288,\,1022] & [576,\,8639] \\
\end{longtable}
\endgroup

\end{landscape}

\section{Moduli-independent flavor symmetries}
\label{app:imagegroups}

\Cref{tab:results331} lists the classical  parent flavor groups 
$\Gshift=\Abhat{S}\rtimes_{\phi^\vee}T_P$ and
$\Ggeo=\Abhat{S}\rtimes_{\phi^\vee}\Outgeo(S)$ of the moduli-independent flavor symmetries. The group acting faithfully on the
classical localization data is instead the image $\operatorname{Im}\rho$
of the monomial representation of \Cref{sec:algorithm-rho}.
Of the 331 geometries, $\Gshift$ is faithful in 254 and $\Ggeo$ in
130. \Cref{tab:imagegroups} therefore lists the 201 geometries for
which at least one parent has a nontrivial kernel; a checkmark denotes a
faithful parent, and the last two columns give the corresponding indices,
i.e. the parent order divided by the image order.

The most common pattern is
\[
[\Gshift:\operatorname{Im}\rho]=1\,,\qquad
[\Ggeo:\operatorname{Im}\rho]=2\,,
\]
which occurs 98 times. In this case, \Cref{eq:imrho} implies
\[
Q=\Ab(S)\,,\qquad K_T=1\,,\qquad |K_{\rm geo}|=2 \,.
\]
Thus, every Abelian charge and every fractional translation is visible on
the localization labels, while one order-two geometric transformation is
not. Since a normal subgroup of order two is central, this invisible
element lies in $Z(\Outgeo)$ and acts trivially on $\Ab(S)$.
If in addition $|\Outlin|=2$, then it complements $T_P$ and
\begin{equation}
\label{eq:directfactor}
\Ggeo\cong\Gshift\x\Z2 \,.
\end{equation}
This occurs in 43 of the 98 rows; whenever both SmallGroup identifiers are
available, the direct-product relation is verified in GAP. For larger
$\Outlin$, the invisible kernel does not need to split off as a direct factor.

The extreme entry $([\Gshift\!:\!\mathrm{Im}\,\rho],[\Ggeo\!:\!\mathrm{Im}\,\rho])=(512,512)$ corresponding to $D_8\text-(1,6)$ is the unique geometry with
$\Sigma_{\mathcal L}=\varnothing$. Its localization representation is
trivial, so both indices simply reproduce the parent orders. The other
rows describe ordinary kernels of the phase or geometric actions on
nonempty localization data.
\begingroup
\small
\setlength{\tabcolsep}{3pt}
\begin{longtable}{
>{\raggedright\arraybackslash}p{0.245\textwidth}
>{\centering\arraybackslash}p{0.050\textwidth}
>{\centering\arraybackslash}p{0.150\textwidth}
>{\centering\arraybackslash}p{0.150\textwidth}
>{\centering\arraybackslash}p{0.112\textwidth}
>{\centering\arraybackslash}p{0.112\textwidth}}
\caption{The $201$ geometries for which at least one classical parent fails to act faithfully on
the classical localization data.  A checkmark means that parent is faithful, so its group is the one
already given in \Cref{tab:results331}; otherwise the GAP data of $\mathrm{Im}\,\rho$ is
listed.  Question marks again mean that no verified identification is available in the GAP/SmallGroups
installation used for the computation.}
\label{tab:imagegroups}\\
\toprule
Geometry & $|\mathcal L|$ & $\mathrm{Im}\,\rho|_{\Gshift}$ & $\mathrm{Im}\,\rho|_{\Ggeo}$ & $[\Gshift\!:\!\mathrm{Im}\,\rho]$ & $[\Ggeo\!:\!\mathrm{Im}\,\rho]$\\
\midrule
\endfirsthead
\toprule
geometry & $|\mathcal L|$ & $\mathrm{Im}\,\rho|_{\Gshift}$ & $\mathrm{Im}\,\rho|_{\Ggeo}$ & $[\Gshift\!:\!\mathrm{Im}\,\rho]$ & $[\Ggeo\!:\!\mathrm{Im}\,\rho]$\\
\midrule
\endhead
\bottomrule
\endfoot
$S_3\text{-}(1,1)$ & 13 & \checkmark & [576,\,8639] & $1$ & $2$ \\
$S_3\text{-}(1,2)$ & 4 & [32,\,49] & [32,\,49] & $9$ & $18$ \\
$S_3\text{-}(2,1)$ & 13 & \checkmark & [576,\,8639] & $1$ & $2$ \\
$S_3\text{-}(2,2)$ & 4 & [32,\,49] & [32,\,49] & $9$ & $18$ \\
$S_3\text{-}(3,2)$ & 4 & [32,\,49] & [32,\,49] & $3$ & $3$ \\
$S_3\text{-}(4,1)$ & 13 & \checkmark & [576,\,8639] & $1$ & $2$ \\
$S_3\text{-}(4,2)$ & 4 & [32,\,49] & [32,\,49] & $9$ & $18$ \\
$S_3\text{-}(5,2)$ & 4 & [32,\,49] & [32,\,49] & $3$ & $3$ \\
$S_3\text{-}(6,1)$ & 5 & \checkmark & [32,\,49] & $1$ & $2$ \\
\addlinespace[1.5pt]
$D_8\text{-}(1,1)$ & 29 & \checkmark & [1024,\,?] & $1$ & $2$ \\
$D_8\text{-}(1,2)$ & 19 & \checkmark & [1024,\,?] & $1$ & $2$ \\
$D_8\text{-}(1,3)$ & 9 & [32,\,49] & [32,\,49] & $32$ & $64$ \\
$D_8\text{-}(1,4)$ & 8 & [256,\,55999] & [256,\,55999] & $2$ & $2$ \\
$D_8\text{-}(1,5)$ & 4 & [32,\,49] & [32,\,49] & $16$ & $16$ \\
$D_8\text{-}(1,6)$ & 0 & [1,\,1] & [1,\,1] & $512$ & $512$ \\
$D_8\text{-}(1,8)$ & 15 & [512,\,?] & [512,\,?] & $2$ & $4$ \\
$D_8\text{-}(1,9)$ & 5 & [16,\,14] & [16,\,14] & $64$ & $128$ \\
$D_8\text{-}(2,1)$ & 20 & \checkmark & [256,\,55683] & $1$ & $2$ \\
$D_8\text{-}(2,2)$ & 11 & [128,\,2194] & [256,\,55999] & $2$ & $2$ \\
$D_8\text{-}(2,3)$ & 10 & [128,\,2194] & [128,\,2194] & $2$ & $4$ \\
$D_8\text{-}(2,4)$ & 7 & [64,\,261] & [128,\,2323] & $4$ & $4$ \\
$D_8\text{-}(2,5)$ & 6 & [64,\,261] & [64,\,261] & $2$ & $2$ \\
$D_8\text{-}(2,6)$ & 2 & [4,\,2] & [4,\,2] & $32$ & $32$ \\
$D_8\text{-}(2,8)$ & 8 & [32,\,46] & [32,\,46] & $8$ & $16$ \\
$D_8\text{-}(3,1)$ & 15 & [128,\,2216] & [128,\,2216] & $2$ & $4$ \\
$D_8\text{-}(3,2)$ & 12 & [128,\,2216] & [256,\,55999] & $2$ & $2$ \\
$D_8\text{-}(3,3)$ & 4 & [32,\,27] & [32,\,27] & $4$ & $4$ \\
$D_8\text{-}(3,4)$ & 13 & [128,\,2216] & [256,\,55805] & $2$ & $2$ \\
$D_8\text{-}(4,1)$ & 13 & \checkmark & [64,\,267] & $1$ & $2$ \\
$D_8\text{-}(4,2)$ & 8 & [32,\,51] & [64,\,261] & $2$ & $2$ \\
$D_8\text{-}(4,3)$ & 5 & [16,\,14] & [32,\,27] & $4$ & $4$ \\
$D_8\text{-}(5,1)$ & 34 & \checkmark & [1024,\,?] & $1$ & $2$ \\
$D_8\text{-}(5,2)$ & 14 & \checkmark & [1024,\,?] & $1$ & $2$ \\
$D_8\text{-}(5,3)$ & 10 & [256,\,55969] & [256,\,55969] & $2$ & $2$ \\
$D_8\text{-}(5,4)$ & 2 & [8,\,3] & [8,\,3] & $64$ & $64$ \\
$D_8\text{-}(5,6)$ & 10 & [256,\,56086] & [256,\,56086] & $4$ & $8$ \\
$D_8\text{-}(6,1)$ & 20 & \checkmark & [256,\,55683] & $1$ & $2$ \\
$D_8\text{-}(6,2)$ & 14 & [128,\,2194] & [256,\,55969] & $2$ & $2$ \\
$D_8\text{-}(6,3)$ & 10 & [128,\,2194] & [128,\,2194] & $2$ & $4$ \\
$D_8\text{-}(6,4)$ & 4 & [32,\,46] & [32,\,46] & $8$ & $16$ \\
$D_8\text{-}(6,5)$ & 6 & [64,\,261] & [64,\,261] & $2$ & $2$ \\
$D_8\text{-}(6,6)$ & 2 & [4,\,2] & [4,\,2] & $32$ & $32$ \\
$D_8\text{-}(6,8)$ & 8 & [32,\,46] & [32,\,46] & $8$ & $16$ \\
$D_8\text{-}(7,1)$ & 13 & \checkmark & [64,\,267] & $1$ & $2$ \\
$D_8\text{-}(7,2)$ & 8 & [32,\,51] & [64,\,261] & $2$ & $2$ \\
$D_8\text{-}(8,1)$ & 19 & \checkmark & [256,\,56086] & $1$ & $2$ \\
$D_8\text{-}(8,2)$ & 4 & [16,\,11] & [16,\,11] & $8$ & $8$ \\
$D_8\text{-}(9,1)$ & 15 & [128,\,2216] & [128,\,2216] & $2$ & $4$ \\
$D_8\text{-}(9,2)$ & 4 & [32,\,27] & [32,\,27] & $4$ & $4$ \\
$D_8\text{-}(9,3)$ & 13 & [128,\,2216] & [256,\,55805] & $2$ & $2$ \\
$D_8\text{-}(9,4)$ & 6 & [64,\,261] & [64,\,261] & $2$ & $2$ \\
\addlinespace[1.5pt]
$A_4\text{-}(1,1)$ & 5 & \checkmark & [12,\,5] & $1$ & $2$ \\
$A_4\text{-}(2,1)$ & 10 & [24,\,15] & [48,\,51] & $2$ & $4$ \\
$A_4\text{-}(2,2)$ & 2 & [3,\,1] & [6,\,1] & $8$ & $16$ \\
$A_4\text{-}(2,3)$ & 6 & \checkmark & [48,\,51] & $1$ & $2$ \\
$A_4\text{-}(3,1)$ & 6 & \checkmark & [12,\,5] & $1$ & $2$ \\
$A_4\text{-}(4,1)$ & 6 & [12,\,5] & [24,\,14] & $4$ & $8$ \\
$A_4\text{-}(4,2)$ & 4 & [12,\,5] & [24,\,14] & $2$ & $4$ \\
$A_4\text{-}(5,1)$ & 6 & [12,\,5] & [24,\,14] & $4$ & $8$ \\
$A_4\text{-}(5,2)$ & 4 & [12,\,5] & [24,\,14] & $2$ & $4$ \\
$A_4\text{-}(6,1)$ & 18 & \checkmark & [96,\,230] & $1$ & $2$ \\
$A_4\text{-}(6,2)$ & 2 & [3,\,1] & [6,\,1] & $8$ & $16$ \\
$A_4\text{-}(7,1)$ & 10 & [24,\,15] & [48,\,51] & $2$ & $4$ \\
$A_4\text{-}(7,2)$ & 2 & [3,\,1] & [6,\,1] & $8$ & $16$ \\
$A_4\text{-}(8,1)$ & 6 & \checkmark & [12,\,5] & $1$ & $2$ \\
$A_4\text{-}(9,1)$ & 6 & [12,\,5] & [24,\,14] & $4$ & $8$ \\
\addlinespace[1.5pt]
$D_{12}\text{-}(1,1)$ & 19 & \checkmark & [64,\,264] & $1$ & $2$ \\
$D_{12}\text{-}(1,2)$ & 4 & [16,\,11] & [16,\,11] & $2$ & $2$ \\
$D_{12}\text{-}(1,4)$ & 9 & [16,\,11] & [16,\,11] & $2$ & $4$ \\
$D_{12}\text{-}(2,1)$ & 19 & \checkmark & [64,\,264] & $1$ & $2$ \\
$D_{12}\text{-}(2,2)$ & 4 & [16,\,11] & [16,\,11] & $2$ & $2$ \\
$D_{12}\text{-}(2,4)$ & 9 & [16,\,11] & [16,\,11] & $2$ & $4$ \\
\addlinespace[1.5pt]
$\Z8\rtimes\Z2\text{-}(1,1)$ & 53 & \checkmark & [512,\,?] & $1$ & $2$ \\
$\Z8\rtimes\Z2\text{-}(1,2)$ & 39 & \checkmark & [512,\,?] & $1$ & $2$ \\
$\Z8\rtimes\Z2\text{-}(1,3)$ & 37 & \checkmark & [512,\,?] & $1$ & $2$ \\
$\Z8\rtimes\Z2\text{-}(2,1)$ & 44 & \checkmark & [512,\,?] & $1$ & $2$ \\
$\Z8\rtimes\Z2\text{-}(4,1)$ & 58 & \checkmark & [512,\,?] & $1$ & $2$ \\
$\Z8\rtimes\Z2\text{-}(4,2)$ & 34 & \checkmark & [256,\,13326] & $1$ & $4$ \\
$\Z8\rtimes\Z2\text{-}(4,4)$ & 32 & \checkmark & [512,\,?] & $1$ & $2$ \\
$\Z8\rtimes\Z2\text{-}(5,1)$ & 44 & \checkmark & [512,\,?] & $1$ & $2$ \\
\addlinespace[1.5pt]
$QD_{16}\text{-}(1,1)$ & 32 & \checkmark & [256,\,55969] & $1$ & $2$ \\
$QD_{16}\text{-}(1,3)$ & 32 & \checkmark & [256,\,55969] & $1$ & $2$ \\
$QD_{16}\text{-}(3,1)$ & 37 & \checkmark & [256,\,56089] & $1$ & $2$ \\
$QD_{16}\text{-}(3,3)$ & 27 & \checkmark & [256,\,56089] & $1$ & $2$ \\
$QD_{16}\text{-}(3,4)$ & 23 & [128,\,2326] & [128,\,2326] & $2$ & $4$ \\
\addlinespace[1.5pt]
$(\Z4\x\Z2)\rtimes\Z2\text{-}(1,1)$ & 75 & \checkmark & [12288,\,?] & $1$ & $2$ \\
$(\Z4\x\Z2)\rtimes\Z2\text{-}(1,2)$ & 43 & \checkmark & [2048,\,?] & $1$ & $2$ \\
$(\Z4\x\Z2)\rtimes\Z2\text{-}(1,3)$ & 42 & \checkmark & [1024,\,?] & $1$ & $2$ \\
$(\Z4\x\Z2)\rtimes\Z2\text{-}(1,4)$ & 24 & \checkmark & [512,\,?] & $1$ & $2$ \\
$(\Z4\x\Z2)\rtimes\Z2\text{-}(1,5)$ & 37 & \checkmark & [2048,\,?] & $1$ & $2$ \\
$(\Z4\x\Z2)\rtimes\Z2\text{-}(1,6)$ & 19 & [256,\,29080] & [256,\,29080] & $2$ & $4$ \\
$(\Z4\x\Z2)\rtimes\Z2\text{-}(1,7)$ & 24 & \checkmark & [512,\,?] & $1$ & $2$ \\
$(\Z4\x\Z2)\rtimes\Z2\text{-}(1,8)$ & 27 & \checkmark & [512,\,?] & $1$ & $2$ \\
$(\Z4\x\Z2)\rtimes\Z2\text{-}(1,11)$ & 33 & [512,\,?] & [1024,\,?] & $2$ & $4$ \\
$(\Z4\x\Z2)\rtimes\Z2\text{-}(1,12)$ & 17 & [128,\,2216] & [128,\,2216] & $4$ & $8$ \\
$(\Z4\x\Z2)\rtimes\Z2\text{-}(1,13)$ & 33 & \checkmark & [3072,\,?] & $1$ & $2$ \\
$(\Z4\x\Z2)\rtimes\Z2\text{-}(1,15)$ & 23 & [256,\,56086] & [512,\,?] & $2$ & $4$ \\
$(\Z4\x\Z2)\rtimes\Z2\text{-}(1,16)$ & 19 & \checkmark & [512,\,?] & $1$ & $2$ \\
$(\Z4\x\Z2)\rtimes\Z2\text{-}(1,18)$ & 19 & [128,\,2216] & [256,\,55751] & $4$ & $8$ \\
$(\Z4\x\Z2)\rtimes\Z2\text{-}(1,19)$ & 17 & [256,\,55805] & [256,\,55805] & $2$ & $4$ \\
$(\Z4\x\Z2)\rtimes\Z2\text{-}(1,20)$ & 18 & \checkmark & [1536,\,?] & $1$ & $2$ \\
$(\Z4\x\Z2)\rtimes\Z2\text{-}(1,21)$ & 14 & [128,\,2216] & [128,\,2216] & $4$ & $8$ \\
$(\Z4\x\Z2)\rtimes\Z2\text{-}(1,22)$ & 12 & [32,\,49] & [32,\,49] & $16$ & $32$ \\
$(\Z4\x\Z2)\rtimes\Z2\text{-}(2,1)$ & 45 & \checkmark & [512,\,?] & $1$ & $2$ \\
$(\Z4\x\Z2)\rtimes\Z2\text{-}(2,3)$ & 29 & \checkmark & [256,\,55643] & $1$ & $2$ \\
$(\Z4\x\Z2)\rtimes\Z2\text{-}(2,5)$ & 18 & \checkmark & [128,\,2194] & $1$ & $2$ \\
$(\Z4\x\Z2)\rtimes\Z2\text{-}(3,1)$ & 30 & \checkmark & [128,\,2328] & $1$ & $2$ \\
$(\Z4\x\Z2)\rtimes\Z2\text{-}(3,4)$ & 30 & \checkmark & [128,\,2328] & $1$ & $2$ \\
$(\Z4\x\Z2)\rtimes\Z2\text{-}(4,1)$ & 61 & \checkmark & [1024,\,?] & $1$ & $2$ \\
$(\Z4\x\Z2)\rtimes\Z2\text{-}(4,2)$ & 37 & \checkmark & [1024,\,?] & $1$ & $2$ \\
$(\Z4\x\Z2)\rtimes\Z2\text{-}(4,3)$ & 43 & \checkmark & [512,\,?] & $1$ & $2$ \\
$(\Z4\x\Z2)\rtimes\Z2\text{-}(4,4)$ & 19 & [64,\,261] & [128,\,2216] & $4$ & $8$ \\
$(\Z4\x\Z2)\rtimes\Z2\text{-}(4,6)$ & 26 & \checkmark & [256,\,56083] & $1$ & $2$ \\
$(\Z4\x\Z2)\rtimes\Z2\text{-}(4,8)$ & 17 & \checkmark & [128,\,2328] & $1$ & $2$ \\
$(\Z4\x\Z2)\rtimes\Z2\text{-}(5,1)$ & 45 & \checkmark & [512,\,?] & $1$ & $2$ \\
$(\Z4\x\Z2)\rtimes\Z2\text{-}(5,3)$ & 31 & [128,\,2320] & [256,\,55683] & $2$ & $4$ \\
$(\Z4\x\Z2)\rtimes\Z2\text{-}(5,5)$ & 26 & \checkmark & [256,\,56083] & $1$ & $2$ \\
$(\Z4\x\Z2)\rtimes\Z2\text{-}(5,7)$ & 17 & \checkmark & [128,\,2328] & $1$ & $2$ \\
$(\Z4\x\Z2)\rtimes\Z2\text{-}(5,10)$ & 19 & [64,\,261] & [128,\,2216] & $4$ & $8$ \\
$(\Z4\x\Z2)\rtimes\Z2\text{-}(5,11)$ & 29 & \checkmark & [256,\,55643] & $1$ & $2$ \\
$(\Z4\x\Z2)\rtimes\Z2\text{-}(5,12)$ & 17 & \checkmark & [128,\,2328] & $1$ & $2$ \\
$(\Z4\x\Z2)\rtimes\Z2\text{-}(5,14)$ & 17 & [64,\,267] & [128,\,1578] & $2$ & $4$ \\
\addlinespace[1.5pt]
$\Z3\x S_3\text{-}(1,1)$ & 65 & \checkmark & [216,\,113] & $1$ & $3$ \\
$\Z3\x S_3\text{-}(1,2)$ & 23 & \checkmark & [54,\,10] & $1$ & $6$ \\
$\Z3\x S_3\text{-}(1,3)$ & 26 & \checkmark & [108,\,28] & $1$ & $3$ \\
$\Z3\x S_3\text{-}(1,4)$ & 20 & \checkmark & [54,\,10] & $1$ & $3$ \\
$\Z3\x S_3\text{-}(3,1)$ & 37 & \checkmark & [72,\,48] & $1$ & $3$ \\
$\Z3\x S_3\text{-}(3,2)$ & 23 & \checkmark & [18,\,5] & $1$ & $6$ \\
$\Z3\x S_3\text{-}(3,3)$ & 22 & \checkmark & [36,\,12] & $1$ & $3$ \\
$\Z3\x S_3\text{-}(3,4)$ & 20 & \checkmark & [18,\,5] & $1$ & $3$ \\
$\Z3\x S_3\text{-}(6,1)$ & 37 & \checkmark & [72,\,48] & $1$ & $3$ \\
$\Z3\x S_3\text{-}(6,2)$ & 22 & \checkmark & [36,\,12] & $1$ & $3$ \\
\addlinespace[1.5pt]
$\Z3\rtimes\Z8\text{-}(1,1)$ & 43 & \checkmark & [64,\,248] & $1$ & $2$ \\
\addlinespace[1.5pt]
$\SL{2,3}\text{\text{-}I}\text{-}(1,1)$ & 39 & \checkmark & [648,\,592] & $1$ & $3$ \\
$\SL{2,3}\text{\text{-}I}\text{-}(2,1)$ & 39 & \checkmark & [216,\,113] & $1$ & $3$ \\
$\SL{2,3}\text{\text{-}I}\text{-}(3,1)$ & 39 & \checkmark & [648,\,592] & $1$ & $3$ \\
\addlinespace[1.5pt]
$\Z4\x S_3\text{-}(1,1)$ & 46 & \checkmark & [128,\,2161] & $1$ & $2$ \\
$\Z4\x S_3\text{-}(1,2)$ & 28 & \checkmark & [64,\,196] & $1$ & $2$ \\
\addlinespace[1.5pt]
$(\Z6\x\Z2)\rtimes\Z2\text{-}(1,1)$ & 37 & \checkmark & [64,\,264] & $1$ & $2$ \\
$(\Z6\x\Z2)\rtimes\Z2\text{-}(2,1)$ & 42 & \checkmark & [192,\,1534] & $1$ & $2$ \\
$(\Z6\x\Z2)\rtimes\Z2\text{-}(2,2)$ & 22 & \checkmark & [32,\,46] & $1$ & $3$ \\
\addlinespace[1.5pt]
$\Z3\x Q_8\text{-}(1,1)$ & 45 & \checkmark & [648,\,592] & $1$ & $3$ \\
\addlinespace[1.5pt]
$S_4\text{-}(1,1)$ & 19 & \checkmark & [32,\,49] & $1$ & $2$ \\
$S_4\text{-}(1,2)$ & 3 & [8,\,3] & [8,\,3] & $2$ & $2$ \\
$S_4\text{-}(2,1)$ & 13 & \checkmark & [16,\,11] & $1$ & $2$ \\
$S_4\text{-}(2,3)$ & 3 & [4,\,2] & [4,\,2] & $2$ & $2$ \\
$S_4\text{-}(3,1)$ & 13 & \checkmark & [16,\,11] & $1$ & $2$ \\
$S_4\text{-}(3,2)$ & 7 & [8,\,3] & [16,\,11] & $2$ & $2$ \\
$S_4\text{-}(3,3)$ & 3 & [4,\,2] & [4,\,2] & $2$ & $2$ \\
$S_4\text{-}(4,1)$ & 10 & \checkmark & [8,\,5] & $1$ & $2$ \\
$S_4\text{-}(5,1)$ & 10 & \checkmark & [8,\,5] & $1$ & $2$ \\
$S_4\text{-}(5,2)$ & 5 & [4,\,2] & [4,\,2] & $2$ & $4$ \\
$S_4\text{-}(5,4)$ & 4 & [4,\,2] & [8,\,3] & $2$ & $2$ \\
$S_4\text{-}(6,1)$ & 10 & \checkmark & [8,\,5] & $1$ & $2$ \\
$S_4\text{-}(6,2)$ & 5 & [4,\,2] & [4,\,2] & $2$ & $4$ \\
\addlinespace[1.5pt]
$\Delta(27)\text{-}(1,3)$ & 16 & [27,\,5] & [54,\,13] & $3$ & $9$ \\
$\Delta(27)\text{-}(1,4)$ & 6 & [3,\,1] & [6,\,1] & $9$ & $36$ \\
$\Delta(27)\text{-}(2,2)$ & 16 & [27,\,5] & [162,\,11] & $3$ & $3$ \\
$\Delta(27)\text{-}(3,3)$ & 16 & [27,\,3] & [54,\,5] & $9$ & $27$ \\
$\Delta(27)\text{-}(3,4)$ & 6 & [3,\,1] & [6,\,1] & $81$ & $324$ \\
\addlinespace[1.5pt]
$(\Z4\x\Z4)\rtimes\Z2\text{-}(1,1)$ & 84 & \checkmark & [256,\,56088] & $1$ & $2$ \\
$(\Z4\x\Z4)\rtimes\Z2\text{-}(1,3)$ & 50 & \checkmark & [256,\,56088] & $1$ & $2$ \\
$(\Z4\x\Z4)\rtimes\Z2\text{-}(1,5)$ & 75 & \checkmark & [256,\,55799] & $1$ & $2$ \\
$(\Z4\x\Z4)\rtimes\Z2\text{-}(1,6)$ & 41 & \checkmark & [256,\,55799] & $1$ & $2$ \\
$(\Z4\x\Z4)\rtimes\Z2\text{-}(2,1)$ & 59 & \checkmark & [128,\,2167] & $1$ & $2$ \\
$(\Z4\x\Z4)\rtimes\Z2\text{-}(2,2)$ & 52 & \checkmark & [128,\,2167] & $1$ & $2$ \\
$(\Z4\x\Z4)\rtimes\Z2\text{-}(2,4)$ & 35 & \checkmark & [128,\,2167] & $1$ & $2$ \\
$(\Z4\x\Z4)\rtimes\Z2\text{-}(3,1)$ & 74 & \checkmark & [256,\,55689] & $1$ & $2$ \\
$(\Z4\x\Z4)\rtimes\Z2\text{-}(3,2)$ & 58 & \checkmark & [256,\,55761] & $1$ & $2$ \\
$(\Z4\x\Z4)\rtimes\Z2\text{-}(3,4)$ & 36 & \checkmark & [128,\,2322] & $1$ & $2$ \\
$(\Z4\x\Z4)\rtimes\Z2\text{-}(3,7)$ & 55 & \checkmark & [128,\,2200] & $1$ & $4$ \\
$(\Z4\x\Z4)\rtimes\Z2\text{-}(3,8)$ & 53 & \checkmark & [256,\,55761] & $1$ & $2$ \\
$(\Z4\x\Z4)\rtimes\Z2\text{-}(3,10)$ & 31 & \checkmark & [64,\,263] & $1$ & $4$ \\
$(\Z4\x\Z4)\rtimes\Z2\text{-}(3,11)$ & 29 & \checkmark & [128,\,2167] & $1$ & $2$ \\
$(\Z4\x\Z4)\rtimes\Z2\text{-}(4,1)$ & 59 & \checkmark & [128,\,2167] & $1$ & $2$ \\
\addlinespace[1.5pt]
$\Z3\x(\Z3\rtimes\Z4)\text{-}(1,1)$ & 73 & \checkmark & [432,\,390] & $1$ & $3$ \\
\addlinespace[1.5pt]
$\Z6\x S_3\text{-}(1,1)$ & 67 & \checkmark & [48,\,45] & $1$ & $3$ \\
$\Z6\x S_3\text{-}(2,1)$ & 53 & \checkmark & [24,\,15] & $1$ & $3$ \\
\addlinespace[1.5pt]
$\Delta(48)\text{-}(1,1)$ & 43 & \checkmark & [24,\,14] & $1$ & $4$ \\
$\Delta(48)\text{-}(1,2)$ & 15 & \checkmark & [24,\,8] & $1$ & $2$ \\
$\Delta(48)\text{-}(2,1)$ & 27 & \checkmark & [48,\,38] & $1$ & $2$ \\
$\Delta(48)\text{-}(2,3)$ & 21 & \checkmark & [24,\,14] & $1$ & $2$ \\
$\Delta(48)\text{-}(3,1)$ & 27 & \checkmark & [48,\,38] & $1$ & $2$ \\
\addlinespace[1.5pt]
$\GL{2,3}\text{-}(1,1)$ & 32 & \checkmark & [32,\,49] & $1$ & $2$ \\
$\GL{2,3}\text{-}(1,4)$ & 32 & \checkmark & [32,\,49] & $1$ & $2$ \\
\addlinespace[1.5pt]
$\SL{2,3}\rtimes\Z2\text{-}(1,1)$ & 55 & \checkmark & [192,\,1524] & $1$ & $2$ \\
$\SL{2,3}\rtimes\Z2\text{-}(1,2)$ & 36 & \checkmark & [96,\,224] & $1$ & $2$ \\
$\SL{2,3}\rtimes\Z2\text{-}(1,3)$ & 41 & \checkmark & [192,\,1524] & $1$ & $2$ \\
\addlinespace[1.5pt]
$\Delta(54)\text{-}(2,3)$ & 19 & \checkmark & [6,\,2] & $1$ & $3$ \\
$\Delta(54)\text{-}(2,4)$ & 14 & \checkmark & [12,\,4] & $1$ & $2$ \\
$\Delta(54)\text{-}(3,3)$ & 19 & \checkmark & [6,\,2] & $1$ & $3$ \\
$\Delta(54)\text{-}(3,4)$ & 14 & \checkmark & [12,\,4] & $1$ & $2$ \\
\addlinespace[1.5pt]
$\Z3\x \SL{2,3}\text{-}(1,1)$ & 73 & \checkmark & [162,\,41] & $1$ & $3$ \\
\addlinespace[1.5pt]
$\Z3\x((\Z6\x\Z2)\rtimes\Z2)\text{-}(1,1)$ & 80 & \checkmark & [24,\,15] & $1$ & $3$ \\
\addlinespace[1.5pt]
$\Delta(96)\text{-}(1,1)$ & 44 & \checkmark & [16,\,14] & $1$ & $2$ \\
$\Delta(96)\text{-}(1,3)$ & 35 & \checkmark & [16,\,11] & $1$ & $2$ \\
$\Delta(96)\text{-}(2,1)$ & 35 & \checkmark & [16,\,11] & $1$ & $2$ \\
$\Delta(96)\text{-}(3,1)$ & 35 & \checkmark & [16,\,11] & $1$ & $2$ \\
$\Delta(96)\text{-}(3,2)$ & 28 & \checkmark & [16,\,11] & $1$ & $2$ \\
\addlinespace[1.5pt]
$\SL{2,3}\rtimes\Z4\text{-}(1,1)$ & 64 & \checkmark & [32,\,48] & $1$ & $2$ \\
$\SL{2,3}\rtimes\Z4\text{-}(1,2)$ & 40 & \checkmark & [32,\,48] & $1$ & $2$ \\
\end{longtable}
\endgroup
\section{Relation to the Narain-space-group normalizer}
\label{app:Narain}
This appendix makes precise the limited Narain statement used in the main
text. We suppress the gauge lattice and Wilson lines and compare only with
the toroidal Narain space group. A lift to a particular heterotic gauge
embedding is an additional, model-dependent step.

\subsection{Geometric and dual Narain transformations}
We use the standard momentum--winding lattice basis of the toroidal Narain
construction \cite{Narain:1985jj,Narain:1986am,GrootNibbelink:2017usl}.
For a symmetric geometric twist $\widehat\vartheta$ and a geometric
linear transformation $\widehat\sigma$, their purely geometric Narain
lifts are
\begin{equation}
\widehat\Theta_\vartheta=
\begin{pmatrix}
\widehat\vartheta&0\\
0&\widehat\vartheta^{-T}
\end{pmatrix},
\qquad
\widehat\Sigma_\sigma=
\begin{pmatrix}
\widehat\sigma&0\\
0&\widehat\sigma^{-T}
\end{pmatrix}.
\label{eq:NarainBlocks}
\end{equation}
Hence, the block-diagonal Narain normalizer condition contains exactly
$\widehat\sigma^{-1}\widehat P\,\widehat\sigma=\widehat P$. For
this purely geometric block, preserving the Narain background is equivalent
to preserving the same $G$ and $B$ used in \Cref{sec:geometric}
\cite{GrootNibbelink:2017usl,Baur:2020jwc}. The finite linear group selected
in the main text is, therefore, the moduli-independent geometric sector of the
toroidal Narain normalizer.

The correspondence is especially simple for translations. Write a toroidal
Narain translation as
\begin{equation}
\widehat T=\binom{t}{u},\qquad t,u\in\RR^6/\ZZ^6 ,
\end{equation}
where $t$ shifts the geometric lattice directions and $u$ the dual
ones. Then, the Narain outer-automorphism condition~\cite{Baur:2020jwc}
\begin{equation}
(\Id-\widehat\Theta_\vartheta)\widehat T\in\ZZ^{12},
\end{equation} separates into
\begin{subequations}\label{eq:NarainSplitTranslations}
\begin{align}
(\Id-\widehat\vartheta)t & \in\ZZ^6\,,\label{eq:tNarain}\\
(\Id-\widehat\vartheta^{-T})u &\in\ZZ^6\,,\label{eq:uNarain}
\end{align}
\end{subequations}
with $\widehat\vartheta\in\widehat P$\,.

On the one hand, \Cref{eq:tNarain} is precisely the definition of $T_P$. Thus, the
fractional translations used in the geometric construction are the
geometric half of the Narain translation sector.
On the other hand, \Cref{eq:uNarain} implies that
\[
n\longmapsto \e^{2\pi\I u^Tn}
\]
is invariant under $n\sim\widehat\vartheta n$, and therefore defines a
character of the coinvariant group $\Lambda/\Lambda_P$. In this sense the
dual Narain translations are the lattice-charge phases. For an untwisted
lattice vertex operator the two halves appear in the familiar combination
\begin{equation}
V_{n,m}\longmapsto
\e^{2\pi\I(m^Tt+n^Tu)}V_{n,m}\,,
\label{eq:NarainVertexPhaseShort}
\end{equation}
where $n$ and $m$ label the two integral momentum--winding lattice
quantum numbers in this convention. Thus, the two translation components are
conjugate to the two Narain lattice directions.

For the full affine space group one must still impose the relations
involving the point-group lifts. Taking the \U1 characters of the finite
Abelian groups in~\Cref{eq:short_exact} gives
\begin{equation}
1\longrightarrow\widehat{\Ab}(P)
\longrightarrow\Abhat{S}
\longrightarrow
\widehat{\frac{\Lambda/\Lambda_P}{\operatorname{Im}d^2}}
\longrightarrow1 \,.
\label{eq:NarainPhaseSequenceShort}
\end{equation}
Here $\widehat{\Ab}(P):=\operatorname{Hom}(\Ab(P),\U1)$ and $\widehat{\frac{\Lambda/\Lambda_P}{\operatorname{Im}d^2}}:=\operatorname{Hom}\left(\frac{\Lambda/\Lambda_P}{\operatorname{Im}d^2},\U1\right)$. 
Only the surviving lattice characters extend to the full space group. Roto-translations can further link these characters to the point-group phases, so that the two need not form independent direct-product factors. These relations are encoded in the full space-group presentation and must be imposed in addition to \Cref{eq:NarainSplitTranslations}.
A \ac{CFT} calculation is required only when one asks for the realization on a
particular physical twisted-state spectrum, not for the classical parent
constructed here.

\subsection{\texorpdfstring{$T^2/\Z2$}{T2/Z2} example}
For $T^2/\Z2$, $\widehat\Theta=-\Id_4$, so the Narain
translation condition gives four independent half-shifts. Two are the
geometric translations of \Cref{sec:Z2example} and generate
$T_P\simeq\Z2^2$; the other two give the independent lattice-charge
phases. Their commutators with the geometric half-shifts supply the central
point-group phase on the twisted labels. The resulting matrices therefore
close to
\begin{equation}
\Gshift=\Ggeo\cong
\frac{D_8\x D_8}{\Z2},
\qquad \mathrm{ID}=[32,49],
\end{equation}
in agreement with the direct geometric construction and the Narain
description \cite{Baur:2020jwc}.

\section{Global common fixed points of all 331 geometries}
\label{app:commonfixed}

For completeness, we record one additional geometric diagnostic, which is not
used to define the flavor representation. For affine point-group
generators
$\widehat g_i=(\widehat\vartheta_i,\widehat\mu_i)$, let us define
\begin{equation}
\begin{aligned}
\widehat{\mathcal F}_{\rm com}(S)
&=\left\{\widehat x\in\RR^6/\ZZ^6\ \middle|\
(\Id-\widehat\vartheta_i)\widehat x-\widehat\mu_i\in\ZZ^6
\ \text{for all }i\right\}\,,\\
\mathcal F_{\rm com}(S)&=e\,\widehat{\mathcal F}_{\rm com}(S),
\qquad
n_{\rm com}=|\widehat{\mathcal F}_{\rm com}(S)|\,.
\end{aligned}
\label{eq:strictCommonFixed}
\end{equation}
This is the set of torus points fixed simultaneously by the complete
affine point-group action. It should not be confused with the
sector-wise localization sets $\mathcal L_{[\vartheta]}$: a common point
is a position in the torus, whereas an element of $\mathcal L$ is a
constructing-element class on which the flavor group acts.

The computation is a single Smith-normal-form problem. We first stack
\begin{equation}
M=
\begin{pmatrix}
\Id-\widehat\vartheta_1\\[-1mm]
\vdots\\
\Id-\widehat\vartheta_r
\end{pmatrix},
\qquad
\widehat\mu=
\begin{pmatrix}
\widehat\mu_1\\[-1mm]
\vdots\\
\widehat\mu_r
\end{pmatrix},
\qquad
M\widehat x\equiv\widehat\mu\bmod{\ZZ^{6r}} \,.
\end{equation}
If $UMV=D$ is a Smith decomposition, the zero rows of $D$ give the
compatibility conditions on $U\widehat\mu$. In the present catalogue
$\operatorname{rank}M=6$ for every geometry. Whenever the system is
compatible,
\begin{equation}
n_{\rm com}=\prod_{j=1}^{6}d_j \,,
\label{eq:ncommonSNF}
\end{equation}
where $d_j$ are the six nonzero Smith invariants; otherwise
$n_{\rm com}=0$.

For a symmorphic representative, $\widehat\mu_i=0$, these are exactly the
same congruences that define $T_P$. Hence
\begin{equation}
n_{\rm com}=|T_P|
\qquad\text{for the symmorphic classes}.
\label{eq:commonTP}
\end{equation}
This equality concerns finite sets, not their physical roles. For example,
$S_3\text-(3,1)$ has $n_{\rm com}=4$ but $4+3$ localization labels, while
$S_3\text-(4,1)$ has $n_{\rm com}=36$ and only $4+9$ such labels. The non-symmorphic example $D_8\text-(9,5)$ is even sharper:
$n_{\rm com}=0$ although $|\mathcal L|=15$.

The exact scan gives
\begin{equation}
\begin{array}{c|rrrrrrrrrrrr}
n_{\rm com}&0&1&2&3&4&6&7&8&9&12&16&36\\
\hline
\#\text{ geometries}&223&28&14&14&30&1&1&11&1&3&4&1
\end{array}.
\label{eq:ncomdistribution}
\end{equation}
The 108 nonempty cases are precisely the symmorphic first affine classes
of the 108 integral lattice ($\mathbb Z$-) classes of the classification;
all 223 remaining affine classes have an
empty strict common fixed set. These numbers describe how the various fixed components intersect and are
useful as auxiliary geometric data, but they do not determine twisted-state
multiplicities or permutation groups.

\clearpage
{\small
\bibliographystyle{OurBibTeX}
\bibliography{Orbifold}

\providecommand{\bysame}{\leavevmode\hbox to3em{\hrulefill}\thinspace}
\begin{thebibliography}{10}

\bibitem{Altarelli:2010gt}
G.~Altarelli and F.~Feruglio, \emph{Discrete flavor symmetries and models of
  neutrino mixing}, Rev. Mod. Phys. \textbf{82} (2010), 2701--2729,
  \texttt{arXiv:1002.0211} [hep-ph].

\bibitem{Ishimori:2010au}
H.~Ishimori, T.~Kobayashi, H.~Ohki, Y.~Shimizu, H.~Okada, and M.~Tanimoto,
  \emph{{Non-Abelian Discrete Symmetries in Particle Physics}}, Prog. Theor.
  Phys. Suppl. \textbf{183} (2010), 1--163, \texttt{arXiv:1003.3552} [hep-th].

\bibitem{Bailin:1999nk}
D.~Bailin and A.~Love, \emph{Orbifold compactifications of string theory},
  Phys. Rept. \textbf{315} (1999), 285--408.

\bibitem{Ramos-Sanchez:2008nwx}
S.~Ramos-S{\'a}nchez, \emph{{Towards Low Energy Physics from the Heterotic
  String}}, Fortsch. Phys. \textbf{10} (2009), 907--1036,
  \texttt{arXiv:0812.3560} [hep-th].

\bibitem{Vaudrevange:2008sm}
P.~K.~S. Vaudrevange, \emph{{Grand Unification in the Heterotic Brane World}},
  Ph.D. thesis, Bonn U., 2008.

\bibitem{Ramos-Sanchez:2024keh}
S.~Ramos-S{\'a}nchez and M.~Ratz, \emph{{Heterotic Orbifold Models}}, {Handbook
  of Quantum Gravity}, Springer, 2024.

\bibitem{Nilles:2012cy}
H.~P. Nilles, M.~Ratz, and P.~K.~S. Vaudrevange, \emph{{Origin of Family
  Symmetries}}, Fortsch. Phys. \textbf{61} (2013), 493--506,
  \texttt{arXiv:1204.2206} [hep-ph].

\bibitem{Nilles:2014owa}
H.~P. Nilles and P.~K.~S. Vaudrevange, \emph{{Geography of Fields in Extra
  Dimensions: String Theory Lessons for Particle Physics}}, Mod. Phys. Lett.
  \textbf{A30} (2015), no.~10, 1530008, \texttt{arXiv:1403.1597} [hep-th].

\bibitem{Hamidi:1986vh}
S.~Hamidi and C.~Vafa, \emph{{Interactions on Orbifolds}}, Nucl. Phys. B
  \textbf{279} (1987), 465--513.

\bibitem{Dixon:1986qv}
L.~J. Dixon, D.~Friedan, E.~J. Martinec, and S.~H. Shenker, \emph{{The
  Conformal Field Theory of Orbifolds}}, Nucl. Phys. \textbf{B282} (1987),
  13--73.

\bibitem{Dixon:1985jw}
L.~J. Dixon, J.~A. Harvey, C.~Vafa, and E.~Witten, \emph{{Strings on
  Orbifolds}}, Nucl. Phys. B \textbf{261} (1985), 678--686.

\bibitem{Dixon:1986jc}
L.~J. Dixon, J.~A. Harvey, C.~Vafa, and E.~Witten, \emph{{Strings on Orbifolds.
  2.}}, Nucl. Phys. B \textbf{274} (1986), 285--314.

\bibitem{Lauer:1989ax}
J.~Lauer, J.~Mas, and H.~P. Nilles, \emph{{Duality and the Role of
  Nonperturbative Effects on the World Sheet}}, Phys. Lett. \textbf{B226}
  (1989), 251--256.

\bibitem{Lauer:1990tm}
J.~Lauer, J.~Mas, and H.~P. Nilles, \emph{{Twisted sector representations of
  discrete background symmetries for two-dimensional orbifolds}}, Nucl. Phys.
  \textbf{B351} (1991), 353--424.

\bibitem{Cremades:2004wa}
D.~Cremades, L.~E. Ib{\'a}{\~n}ez, and F.~Marchesano, \emph{{Computing Yukawa
  couplings from magnetized extra dimensions}}, JHEP \textbf{05} (2004), 079,
  \texttt{hep-th/0404229}.

\bibitem{Buchbinder:2016yuh}
E.~I. Buchbinder, A.~Constantin, J.~Gray, and A.~Lukas, \emph{{Yukawa
  Unification in Heterotic String Theory}}, Phys. Rev. D \textbf{94} (2016),
  no.~4, 046005, \texttt{arXiv:1606.04032} [hep-th].

\bibitem{Candelas:2017ive}
P.~Candelas and C.~Mishra, \emph{{Highly Symmetric Quintic Quotients}},
  Fortsch. Phys. \textbf{66} (2018), no.~4, 1800017, \texttt{arXiv:1709.01081}
  [hep-th].

\bibitem{Lukas:2017vqp}
A.~Lukas and C.~Mishra, \emph{{Discrete Symmetries of Complete Intersection
  Calabi-Yau Manifolds}}, Commun. Math. Phys. \textbf{379} (2020), 847--865,
  \texttt{arXiv:1708.08943} [hep-th].

\bibitem{Braun:2017cys}
A.~P. Braun, A.~Lukas, and C.~Sun, \emph{{Discrete Symmetries of Calabi-Yau
  Hypersurfaces in Toric Four-Folds}}, Commun. Math. Phys. \textbf{360} (2018),
  935--984, \texttt{arXiv:1704.07812} [hep-th].

\bibitem{Gray:2021fcq}
J.~Gray and J.~Wang, \emph{{Free Quotients of Favorable Calabi-Yau Manifolds}},
  JHEP \textbf{07} (2022), 116, \texttt{arXiv:2112.12683} [hep-th].

\bibitem{Ishiguro:2021ccl}
K.~Ishiguro, T.~Kobayashi, and H.~Otsuka, \emph{{Symplectic modular symmetry in
  heterotic string vacua: flavor, CP, and R-symmetries}}, JHEP \textbf{01}
  (2022), 020, \texttt{arXiv:2107.00487} [hep-th].

\bibitem{Almumin:2021fbk}
Y.~Almumin, M.-C. Chen, V.~Knapp-P\'erez, S.~Ramos-S\'anchez, M.~Ratz, and
  S.~Shukla, \emph{{Metaplectic Flavor Symmetries from Magnetized Tori}}, JHEP
  \textbf{05} (2021), 078, \texttt{arXiv:2102.11286} [hep-th].

\bibitem{Ishiguro:2024mhy}
K.~Ishiguro, T.~Kobayashi, S.~Nishimura, and H.~Otsuka, \emph{{Modular forms
  and hierarchical Yukawa couplings in heterotic Calabi-Yau
  compactifications}}, JHEP \textbf{08} (2024), 088, \texttt{arXiv:2402.13563}
  [hep-th].

\bibitem{Kobayashi:2024fsm}
T.~Kobayashi, K.~Nasu, R.~Nishida, H.~Otsuka, and S.~Takada, \emph{{Flavor
  symmetries from modular subgroups in magnetized compactifications}}, JHEP
  \textbf{12} (2024), 128, \texttt{arXiv:2409.02458} [hep-th].

\bibitem{Dong:2025csr}
J.~Dong, T.~Kobayashi, R.~Nishida, S.~Nishimura, and H.~Otsuka, \emph{{Coupling
  Selection Rules in Heterotic Calabi-Yau Compactifications}}, JHEP \textbf{09}
  (2025), 012, \texttt{arXiv:2504.09773} [hep-th].

\bibitem{Dong:2026yte}
J.~Dong, T.~Kobayashi, S.~Miyamoto, and H.~Otsuka, \emph{{Yukawa textures with
  enhanced symmetries in heterotic Calabi-Yau compactifications}}, JHEP
  \textbf{06} (2026), 273, \texttt{arXiv:2603.00864} [hep-th].

\bibitem{Adulpravitchai:2009xyz}
A.~Adulpravitchai, A.~Blum, and M.~Lindner, \emph{{Non-Abelian Discrete Flavor
  Symmetries from $T^2/Z_N$ Orbifolds}}, JHEP \textbf{07} (2009), 053,
  \texttt{arXiv:0906.0468} [hep-ph].

\bibitem{Carballo-Perez:2016ooy}
B.~Carballo-P{\'e}rez, E.~Peinado, and S.~Ramos-S{\'a}nchez,
  \emph{{$\Delta(54)$ flavor phenomenology and strings}}, JHEP \textbf{12}
  (2016), 131, \texttt{arXiv:1607.06812} [hep-ph].

\bibitem{Beye:2014nxa}
F.~Beye, T.~Kobayashi, and S.~Kuwakino, \emph{{Gauge Origin of Discrete Flavor
  Symmetries in Heterotic Orbifolds}}, Phys. Lett. \textbf{B736} (2014),
  433--437, \texttt{arXiv:1406.4660} [hep-th].

\bibitem{Kobayashi:2004ya}
T.~Kobayashi, S.~Raby, and R.-J. Zhang, \emph{{Searching for realistic 4d
  string models with a Pati-Salam symmetry: Orbifold grand unified theories
  from heterotic string compactification on a Z(6) orbifold}}, Nucl. Phys. B
  \textbf{704} (2005), 3--55, \texttt{hep-ph/0409098}.

\bibitem{Ko:2007dz}
P.~Ko, T.~Kobayashi, J.-h. Park, and S.~Raby, \emph{String-derived {D4} flavor
  symmetry and phenomenological implications}, Phys. Rev. \textbf{D76} (2007),
  035005, \texttt{arXiv:0704.2807 [hep-ph]}.

\bibitem{Kobayashi:2006wq}
T.~Kobayashi, H.~P. Nilles, F.~Pl{\"o}ger, S.~Raby, and M.~Ratz, \emph{{Stringy
  origin of non-Abelian discrete flavor symmetries}}, Nucl. Phys. B
  \textbf{768} (2007), 135--156, \texttt{hep-ph/0611020}.

\bibitem{Olguin-Trejo:2018wpw}
Y.~Olgu{\'i}n-Trejo, R.~P{\'e}rez-Mart{\'i}nez, and S.~Ramos-S{\'a}nchez,
  \emph{{Charting the flavor landscape of MSSM-like Abelian heterotic
  orbifolds}}, Phys. Rev. \textbf{D98} (2018), no.~10, 106020,
  \texttt{arXiv:1808.06622} [hep-th].

\bibitem{Ramos-Sanchez:2018edc}
S.~Ramos-S{\'a}nchez and P.~K. Vaudrevange, \emph{{Note on the space group
  selection rule for closed strings on orbifolds}}, JHEP \textbf{01} (2019),
  055, \texttt{arXiv:1811.00580} [hep-th].

\bibitem{Narain:1985jj}
K.~S. Narain, \emph{{New Heterotic String Theories in Uncompactified Dimensions
  \ensuremath{<} 10}}, Phys. Lett. B \textbf{169} (1986), 41--46.

\bibitem{Narain:1986am}
K.~S. Narain, M.~H. Sarmadi, and E.~Witten, \emph{{A Note on Toroidal
  Compactification of Heterotic String Theory}}, Nucl. Phys. B \textbf{279}
  (1987), 369--379.

\bibitem{GrootNibbelink:2017usl}
S.~Groot~Nibbelink and P.~K.~S. Vaudrevange, \emph{{T-duality orbifolds of
  heterotic Narain compactifications}}, JHEP \textbf{04} (2017), 030,
  \texttt{arXiv:1703.05323} [hep-th].

\bibitem{Baur:2019kwi}
A.~Baur, H.~P. Nilles, A.~Trautner, and P.~K. Vaudrevange, \emph{{Unification
  of Flavor, CP, and Modular Symmetries}}, Phys. Lett. B \textbf{795} (2019),
  7--14, \texttt{arXiv:1901.03251} [hep-th].

\bibitem{Baur:2019iai}
A.~Baur, H.~P. Nilles, A.~Trautner, and P.~K. Vaudrevange, \emph{{A String
  Theory of Flavor and $\mathcal{CP}$}}, Nucl. Phys. B \textbf{947} (2019),
  114737, \texttt{arXiv:1908.00805} [hep-th].

\bibitem{Nilles:2020gvu}
H.~P. Nilles, S.~Ramos-S\'anchez, and P.~K.~S. Vaudrevange, \emph{{Eclectic
  flavor scheme from ten-dimensional string theory - II. Detailed technical
  analysis}}, Nucl. Phys. B \textbf{966} (2021), 115367,
  \texttt{arXiv:2010.13798} [hep-th].

\bibitem{Baur:2020jwc}
A.~Baur, M.~Kade, H.~P. Nilles, S.~Ramos-S{\'a}nchez, and P.~K.~S. Vaudrevange,
  \emph{{The eclectic flavor symmetry of the $\boldsymbol{\mathbb{Z}_2}$
  orbifold}}, JHEP \textbf{02} (2021), 018, \texttt{arXiv:2008.07534} [hep-th].

\bibitem{Baur:2024qzo}
A.~Baur, H.~P. Nilles, S.~Ramos-S{\'a}nchez, A.~Trautner, and P.~K.~S.
  Vaudrevange, \emph{{The eclectic flavor symmetries of $
  \mathbb{T}^2/\mathbb{Z}_K$ orbifolds}}, JHEP \textbf{09} (2024), 159,
  \texttt{arXiv:2405.20378} [hep-th].

\bibitem{Kobayashi:2018rad}
T.~Kobayashi, S.~Nagamoto, S.~Takada, S.~Tamba, and T.~H. Tatsuishi,
  \emph{{Modular symmetry and non-Abelian discrete flavor symmetries in string
  compactification}}, Phys. Rev. \textbf{D97} (2018), no.~11, 116002,
  \texttt{arXiv:1804.06644} [hep-th].

\bibitem{Nilles:2020tdp}
H.~P. Nilles, S.~Ramos-Sánchez, and P.~K.~S. Vaudrevange, \emph{{Eclectic
  flavor scheme from ten-dimensional string theory -- I. Basic results}}, Phys.
  Lett. B \textbf{808} (2020), 135615, \texttt{arXiv:2006.03059} [hep-th].

\bibitem{Nilles:2020nnc}
H.~P. Nilles, S.~Ramos-S{\'a}nchez, and P.~K. Vaudrevange, \emph{{Eclectic
  Flavor Groups}}, JHEP \textbf{02} (2020), 045, \texttt{arXiv:2001.01736}
  [hep-ph].

\bibitem{Baur:2021mtl}
A.~Baur, M.~Kade, H.~P. Nilles, S.~Ramos-S{\'a}nchez, and P.~K.~S. Vaudrevange,
  \emph{{Completing the eclectic flavor scheme of the $\mathbb Z_2$ orbifold}},
  JHEP \textbf{06} (2021), 110, \texttt{arXiv:2104.03981} [hep-th].

\bibitem{Li:2025bsr}
X.~Li, X.-G. Liu, H.~P. Nilles, M.~Ratz, and A.~Stewart, \emph{{Flavor
  symmetries and winding modes}}, JHEP \textbf{09} (2025), 026,
  \texttt{arXiv:2506.12887} [hep-th].

\bibitem{Baykara:2024vss}
Z.~K. Baykara, H.-C. Tarazi, and C.~Vafa, \emph{{The Quasicrystalline String
  Landscape}}, Phys. Rev. D \textbf{111} (2025), no.~8, 086025,
  \texttt{arXiv:2406.00129} [hep-th].

\bibitem{Funakoshi:2025lxs}
S.~Funakoshi, Y.~Koga, and H.~Otsuka, \emph{{Classification of modular
  symmetries in non-supersymmetric heterotic string theories}}, JHEP
  \textbf{08} (2026), 008, \texttt{arXiv:2503.23741} [hep-th].

\bibitem{Fischer:2012qj}
M.~Fischer, M.~Ratz, J.~Torrado, and P.~K.~S. Vaudrevange,
  \emph{{Classification of symmetric toroidal orbifolds}}, JHEP \textbf{01}
  (2013), 084, \texttt{arXiv:1209.3906} [hep-th].

\bibitem{Fischer:2013qza}
M.~Fischer, S.~Ramos-S{\'a}nchez, and P.~K.~S. Vaudrevange, \emph{{Heterotic
  non-Abelian orbifolds}}, JHEP \textbf{07} (2013), 080,
  \texttt{arXiv:1304.7742} [hep-th].

\bibitem{Konopka:2012gy}
S.~J.~H. Konopka, \emph{{Non Abelian orbifold compactifications of the
  heterotic string}}, JHEP \textbf{07} (2013), 023, \texttt{arXiv:1210.5040}
  [hep-th].

\bibitem{Hernandez-Segura:2025sfr}
M.~Hern{\'a}ndez-Segura and S.~Ramos-S{\'a}nchez, \emph{{Non-Abelian orbifolds
  of the SO(32) heterotic string}}, Phys. Rev. D \textbf{112} (2025), no.~6,
  066002, \texttt{arXiv:2506.08370} [hep-th].

\bibitem{Kaidi:2024sgr}
J.~Kaidi, Y.~Tachikawa, and H.~Y. Zhang, \emph{{On a class of selection rules
  without group actions in field theory and string theory}}, SciPost Phys.
  \textbf{17} (2024), no.~6, 169, \texttt{arXiv:2402.00105} [hep-th].

\bibitem{Kobayashi:2025nis}
T.~Kobayashi, R.~Nishida, and H.~Otsuka, \emph{{Non-invertible selection rules
  on heterotic non-Abelian orbifolds}}, JHEP \textbf{03} (2026), 158,
  \texttt{arXiv:2509.10019} [hep-th].

\bibitem{Funakoshi:2024nif}
S.~Funakoshi, T.~Kobayashi, and H.~Otsuka, \emph{{Quantum aspects of
  non-invertible flavor symmetries in intersecting/magnetized D-brane models}},
  JHEP \textbf{04} (2025), 183, \texttt{arXiv:2412.12524} [hep-th].

\bibitem{Knapp-Perez:2025cht}
V.~Knapp-P{\'e}rez, X.-G. Liu, H.~P. Nilles, and S.~Ramos-S{\'a}nchez,
  \emph{{Demystifying stringy miracles with eclectic flavor symmetries}}, Phys.
  Rev. D \textbf{113} (2026), no.~10, 106015, \texttt{arXiv:2512.21382}
  [hep-th].

\bibitem{Liu:2025lym}
X.-G. Liu and M.~Ratz, \emph{{Modular Zeros}}, Phys. Lett. B \textbf{875}
  (2026), 140372, \texttt{arXiv:2512.21386} [hep-th].

\bibitem{Kobayashi:2011cw}
T.~Kobayashi, S.~L. Parameswaran, S.~Ramos-S{\'a}nchez, and I.~Zavala,
  \emph{{Revisiting Coupling Selection Rules in Heterotic Orbifold Models}},
  JHEP \textbf{05} (2012), 008, \texttt{arXiv:1107.2137} [hep-th], [Erratum:
  JHEP12,049(2012)].

\bibitem{Nilles:2013lda}
H.~P. Nilles, S.~Ramos-S{\'a}nchez, M.~Ratz, and P.~K.~S. Vaudrevange, \emph{{A
  note on discrete $R$ symmetries in $\Z{6}$-$\mathrm{II}$ orbifolds with
  Wilson lines}}, Phys. Lett. \textbf{B726} (2013), 876--881,
  \texttt{arXiv:1308.3435} [hep-th].

\bibitem{Ratcliffe:2009}
J.~G. Ratcliffe and S.~T. Tschantz, \emph{Abelianization of space groups}, Acta
  Cryst. A \textbf{65} (2009), no.~1, 18--27,
  \texttt{https://onlinelibrary.wiley.com/doi/abs/10.1107/S0108767308036222}.

\bibitem{Brown:1982}
K.~S. Brown, \emph{Cohomology of groups}, Graduate Texts in Mathematics,
  vol.~87, Springer-Verlag, New York, 1982.

\bibitem{GAP4}
The GAP~Group, \emph{{GAP -- Groups, Algorithms, and Programming, Version
  4.11.1}}, 2021, \texttt{\url{https://www.gap-system.org}}.

\bibitem{SmallGrpManual}
{The GAP Group}, \emph{The small groups library, gap package smallgrp},
  \url{https://docs.gap-system.org/pkg/smallgrp/doc/chap1.html}, 2026.

\bibitem{Banks:2010zn}
T.~Banks and N.~Seiberg, \emph{Symmetries and strings in field theory and
  gravity}, Phys. Rev. D \textbf{83} (2011), 084019, \texttt{arXiv:1011.5120}
  [hep-th].

\bibitem{Harlow:2018jwu}
D.~Harlow and H.~Ooguri, \emph{Symmetries in quantum field theory and quantum
  gravity}, Commun. Math. Phys. \textbf{383} (2021), 1669--1804,
  \texttt{arXiv:1810.05338} [hep-th].

\end{thebibliography}

}

\begin{acronym}
 \acro{CFT}{Conformal Field Theory}
 \acroplural{CFT}[CFTs]{Conformal Field Theories}
\end{acronym}
\end{document}